\documentclass[]{aa}
\usepackage{graphicx,amsmath}
\usepackage{natbib}
\bibpunct{(}{)}{,}{a}{}{,}

\usepackage{txfonts}

\begin{document}

\title{Projection of circumstellar disks on their environments} 

\author{K. M. Pontoppidan \inst{1} \and C. P. Dullemond \inst{2}}

\offprints{K. M. Pontoppidan}
\institute{Leiden Observatory, P.O. Box 9513, NL-2300 RA Leiden, Netherlands\\
                \email{pontoppi@strw.leidenuniv.nl}
	\and
                Max-Planck Institut f{\"u}r Astrophysik, P.O. Box 1317, 85741 Garching, Germany
                 }
             
\date{}

\abstract{
We use a 3D Monte Carlo radiative transfer code to study the projection of large shadows 
by circumstellar disks around young stellar objects on surrounding reflection nebulosity.
It is shown that for a wide range of parameters a small (10-100\,AU) circumstellar disk 
can project a large (1\,000-10\,000\,AU) dark band in the near-infrared that often resembles 
a massive edge-on disk. The disk shadows are divided into two basic types, depending on the 
distribution of the reflecting material and the resulting morphology of the shadows in the 
near-infrared. Two YSOs associated with bipolar nebulosity, CK 3/EC 82 illuminating the Serpens 
Reflection Nebula (SRN) and Ced 110 IRS 4 in the Chamaeleon I molecular cloud, are modelled in 
detail as disk shadows. Spectral energy distributions of the two sources are collected using both 
archival ISO data and new Spitzer-IRS data. An axisymmetric model consisting of a small disk and a 
spherically symmetric envelope can reproduce the near-infrared images and full spectral energy 
distributions of the two disk shadow candidates. It is shown that the model fits can be used to 
constrain the geometry of the central disks due to the magnifying effect of the projection. The 
presence of a disk shadow may break a number of degeneracies encountered when fitting to the SED 
only. Specifically, the inclination, flaring properties and extinction toward the central star may 
be independently determined from near-infrared images of disk shadows. Constraints on the disk 
mass and size can be extracted from a simultaneous fit of SEDs and images. We find that the CK 3 
disk must have a very low mass in opacity-producing, small ($\lesssim 10\,\mu$m) dust grains 
(corresponding to a total mass of $\sim 7\times 10^{-6}\,\rm {M_{\odot}}$, assuming a gas-to-dust ratio of 100) to simultaneously 
reproduce the very strong silicate emission features and the near-infrared edge-on morphology. 
Ced 110 IRS 4 requires that a roughly spherical cavity with radius $\sim 500$\,AU centered on the 
central star-disk system is carved out of the envelope to reproduce the near-infrared images. We 
show that in some cases the bipolar nebulosity created by a disk shadow may resemble the effect of 
a physical bipolar cavity where none exists. We find that a disk unresolved in near-infrared 
images, but casting a large disk shadow, can be modelled at a level of sophistication approaching 
that of an edge-on disk with resolved near-infrared images. Selection criteria are given for 
distinguishing disk shadows from genuine large disks. It is found that the most obvious observable 
difference between a disk shadow and a large optically thick disk is that the disk shadows have a 
compact near-infrared source near the center of the dark band. High resolution imaging and/or 
polarimetry should reveal the compact source in the center of a disk shadow as an edge-on disk.
Finally, it is shown that disk shadows can be used to select edge-on disks suitable for observing 
ices located inside the disk.

}

\maketitle

\section{Introduction}
The strong near- and mid-infrared excess measured from T Tauri and Herbig
Ae/Be stars is generally regarded as strong evidence that these young stars
are still surrounded by remnants of the accretion disks from which they were
formed. Indeed, the infrared spectral signatures from these
disks agree well with those theoretically expected for such `protoplanetary
disks'  \citep{Kenyon87, Adams87, Chiang97,Dominik03}. Recent observations have put constraints of the spatial structure
of the inner regions of disks. For instance, 
interferometric observations of Herbig Ae/Be stars in the near-infrared by \cite{Millan-Gabet01} and 
\cite{Eisner03} have shown that the sources are larger than expected for thermal emission from a disk 
extending all the way to the central star. Although this is possibly due to
gaps in the inner disk or scattering of the near-infrared photons into the line of sight from the disk 
surface, it illustrates the difficulties in unambiguously determining the geometry of circumstellar 
material. So far, perhaps the most compelling evidence for the disk-like nature of circumstellar material 
within a few 100\,AU of a star is given by the optical and near-infrared images
of edge-on disks \citep{Padgett99} and spatially resolved kinematic evidence from molecular line 
observations \citep[e.g.][]{Mannings97}. In these
images, in particular those of Padgett et al., bright bipolar reflection
nebulae feature a conspicuous `dark lane'. This dark lane, which is due to obscuration of the bright inner regions by
the dark outer parts of the disk, is seen whenever
the disk is viewed close to edge-on. If a bright background illumination is
present \cite[e.g.][]{McCaughrean96}, the silhouette of the disk clearly indicates
the extent and shape of the disk itself. However, when the only illuminating source of the circumstellar 
matter is the central star, the disk appears as a dark lane without a clear outer edge. These
images allow one to detemine the geometric thickness and the minimum radius
of such disks.

There may, however, be another possibility to produce a `dark lane'
signature. \cite{Hodapp04} suggested that the bipolar morphology of the young star ASR 41 in
the NGC 1333 molecular cloud is caused by the shadow of a much smaller disk projected on an
envelope reflecting the light from the central star. The general possibility of disks casting shadows much 
larger than their size is also clear from the models by \cite{Whitney03}.

In this paper we discuss the possibility that a circumstellar
disk around a young star may commonly project a large shadow onto the matter
surrounding the system to produce a broader range of morphologies of reflection nebulae than
that already proposed by \cite{Hodapp04}. We have selected two young stars that probably cast disk 
shadows on associated reflection nebulosity; CK 3 of the Serpens reflection nebula and Ced 110 IRS 4 in 
the Chamaeleon molecular cloud, and we model them in detail using a Monte Carlo radiative transfer code.

This article is organised as follows: \S\ref{Scenarios} defines the scenarios which are expected to 
produce disk shadows. \S\ref{Observations_shadow} describes the imaging and spectroscopy for the two 
disk shadow candidates which have been collected to constrain the models. The radiative transfer model 
is treated in \S\ref{RADTRANS}. In \S\ref{CK3} and \S\ref{CED}, the two disk shadow candidates and the 
best fitting radiative transfer models are described. Finally, selection criteria for distinguishing 
between genuine disk shadows and
large massive edge-on disks are discussed in \S\ref{Selection}.

\section{Scenarios}
\label{Scenarios}

In this section, the basic geometry of shadows cast by circumstellar disks is explored. In particular,
we divide disk shadows into two basic types, based on the distribution of circumstellar, reflecting 
material. Quantitative aspects of the two types are discussed in the context of specific objects in 
\S\ref{CK3} and \S\ref{CED}.

\begin{figure}
   \centering
   \includegraphics[width=7.cm]{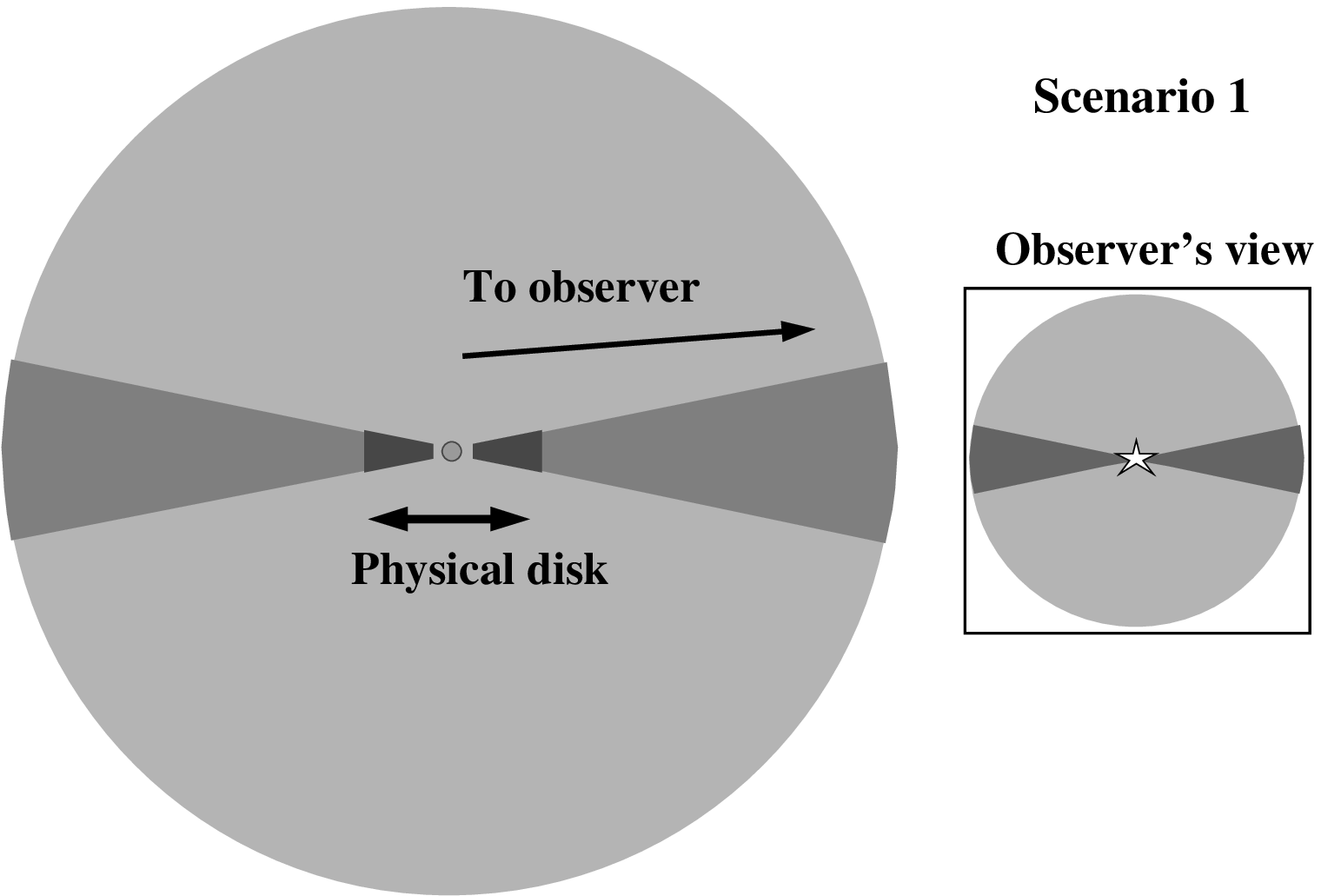}
   \includegraphics[width=7.cm]{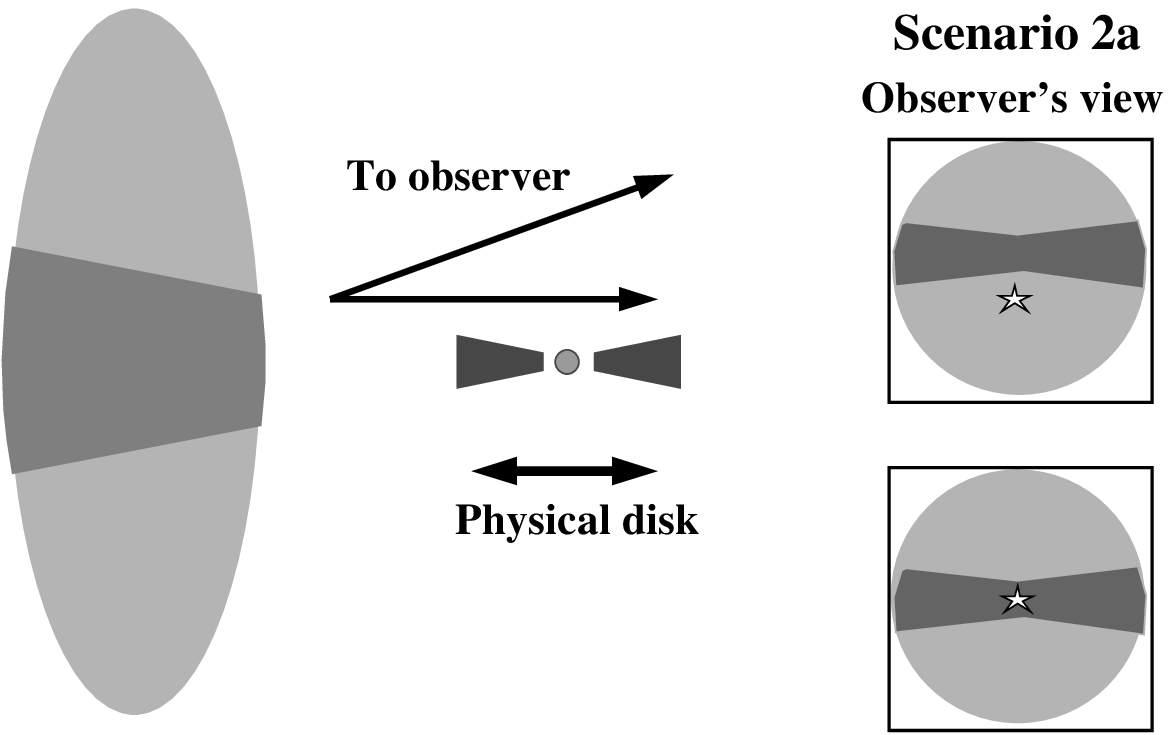}
   \includegraphics[width=7.cm]{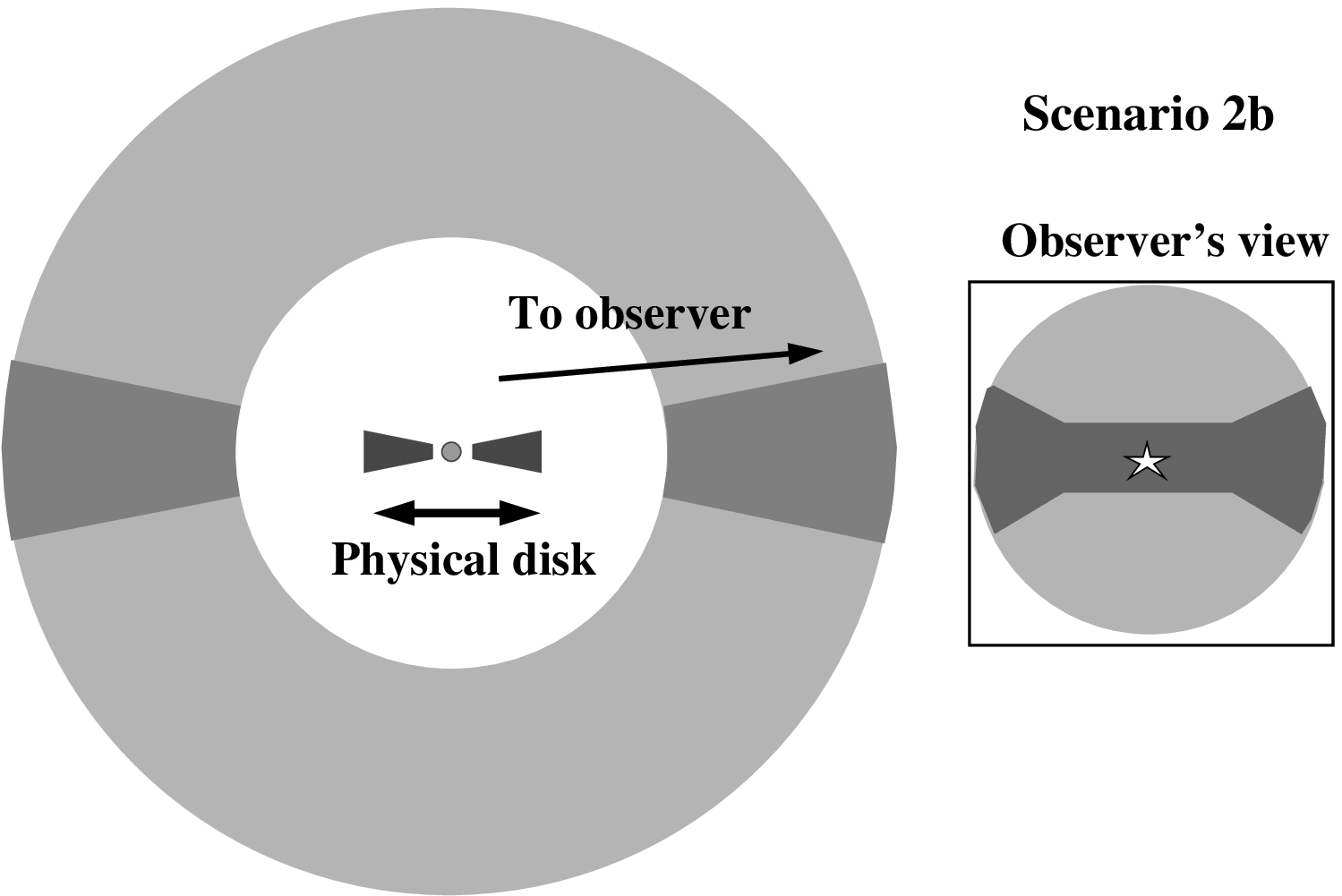}
   \caption{Cartoon illustrating the disk shadows produced by different distributions of circumstellar dust. The distribution
of dust is sketched on the left, while the insets to the right sketches what an observer sees in an optical or near-infrared image.
The star in the insets indicate the location of a compact source, typically created by light scattered in the disk surface.
{\it Top panel:} Scenario 1, in which the disk is surrounded by a volume-filling distribution of dust. This produces a wedge-like
shadow. {\it Middle panel:} Scenario 2a, in which a cloud is located behind the star-disk system. In this case the observer sees
a dark band that may be completely offset from the central star if the star-disk system is not viewed edge-on. An observer's view
for two different inclinations of the star-disk system is shown. 
{\it Bottom panel:} Scenario 2b, in which a central cavity is excavated in an otherwise volume-filling distribution of dust. An observer
will see a broad dark band with the star located in the center of the shadowed band.
   }
   \label{screen_cartoon}
\end{figure}

In principle, an extended shadow, visible to an observer in the optical to near-infrared wavebands, 
should be produced whenever the disk is viewed
at a sufficiently inclined angle ($>60\degr$ measured from the line-of sight to the polar axis of the disk) and enough dusty matter 
is present in an envelope to produce extended scattering nebulosity in the near-infrared. Conversely, the envelope cannot be so 
optically thick to near-infrared photons that none escape to be observed. Another requirement is that the disk mid-plane 
is not optically thin 
to near-infrared photons. The shadow of such a disk may be arbitrarily large, depending only on
the extent of the illuminated nebula and not on the actual physical extent of the disk itself. We divide shadows 
produced by circumstellar disks into two basic categories based on the distribution of extended scattering material and 
therefore on the observable near-infrared morphology. The first type of disk shadow occurs when the star-disk system is 
embedded in a volume-filling distribution of dust. This type of shadow
will create a symmetric bipolar morphology of the reflection nebulosity. The shadow itself will acquire a 
symmetric wedge-like shape, which may resemble the `dark-lane' signature
of edge-on disks seen in absorption. This happens because only photons emitted at angles larger than the opening 
angle of the disk survive to be scattered on the envelope material. The wedge-like shadow produced by a volume-filling
distribution of envelope material is likely the 
most common configuration, and will be referred to as `scenario 1'.  ASR 41 was shown by \cite{Hodapp04} 
to be a likely `scenario 1' candidate. The top panel in Fig. \ref{screen_cartoon} sketches the volume-filling scenario.

Different types of shadowing morphologies can be produced if the scattering dust 
does not fill the volume surrounding the central star-disk system.  A simple scenario with a non-volume-filling
dust distribution may
be that of a disk projecting its shadow onto a nearby optically thick background cloud. For an observer, such shadows
will take shape of a dark band, possibly offset from the star-disk system. In essence, a scattering cloud offset from
the star will act as a `screen' onto which the shadow is projected. This scenario is sketched in the middle panel of
Fig. \ref{screen_cartoon}. We refer to such screen projection shadows as `scenario 2' cases. 
`Scenario 2' shadows may exhibit a range of morphologies, depending on the configuration of the screen. The simplest `scenario 2' shadow
is perhaps that of a cloud located entirely behind the star-disk system. This scenario is sketched in Fig. 1 as `scenario 2a', and is
characterised by a dark, broad shadow with a size depending on both the opening angle of the disk and the distance between the scattering
cloud and the star-disk system. The strongly non-spherical dust distribution is the only configuration allowing the star-disk 
system being offset from the center of the shadowed band, as seen by an observer.  
Another potentially important version of a shadow cast of a non-volume filling dust distribution is that of a spherical envelope
with a central spherical or cylindrical cavity.  In this case, the walls 
of the cavity act as a screen. 
The shape of the dark band will in this case be almost rectangular and the star-disk system itself will 
characteristically be seen as a compact near-infrared source
in the exact center of the dark band. This case of a spherical envelope with a central spherical cavity is sketched in 
the bottom panel of Fig. \ref{screen_cartoon} and marked as `scenario 2b'. 

It is important to note that all disk shadows will contain a compact object at the location of the star-disk system. 
In almost all cases where a shadow is seen, 
the compact emission will be entirely due to light scattered on the surface of the disk. Only in `scenario 2a' can the star be seen 
directly if the inclination is sufficiently low. In this case, the observer will see the star-disk system located outside the shadowed 
band, as shown in the sketch of `scenario 2a' in Fig. \ref{screen_cartoon}.

Clearly, many young stars are illuminating nebulosity with a complex structure. Thus, axisymmetric envelope models 
may not always be sufficient. Such non-volume filling cases are likely common for more evolved young stars, where the
surrounding envelopes have been significantly disturbed and dispersed by outflow activity.

We have identified two new candidates for circumstellar disks that project a shadow on large reflection 
nebulae; one for `scenario 1' and one for `scenario 2b' as illustrated in Fig. \ref{screen_cartoon}. 
The `scenario 1' candidate is the area of the Serpens reflection nebula 
that is illuminated by the star CK 3 (also known as EC 82 or SVS 2). The `scenario 2b' candidate is IRS 4, 
the brightest far-infrared source in the Cederblad 110 cloud in the Chamaeleon star-forming complex.  We 
present detailed Monte Carlo radiative transfer models that can reproduce the near-infrared morphological 
characteristics as well as the spectral energy distributions (SEDs) of the two candidates. Additionally, we show how 
such models aided by the presence of a disk shadow can be used to gain unique insight into the structure 
of circumstellar disks too small to be resolved by imaging. 

\section{Observations}
\label{Observations_shadow}
We use a range of both imaging and spectroscopic data to constrain the models.
Archival $JHK_s$ images from the Infrared Spectrometer And Array Camera (ISAAC) mounted on
UT1-Antu of the Very Large Telescope (VLT) at Cerro Paranal in Chile have been extracted from the 
VLT archive\footnote{Based on observations made with ESO Telescopes at Paranal Observatory under 
programmes\,{63.I-0691}, {164.I-0605} and {67.C-0600}.}. The images were taken during moderate to good 
seeing conditions, resulting in infrared spatial resolutions of $\sim 0\farcs7$.  
One pointing obtained on May 3, 2001 is of the south-eastern core of the Serpens star-forming cloud, 
while the other obtained on April 29, 1999 is of the Cederblad 110 region in the Chamaeleon star-forming 
complex. The ISAAC images were reduced following standard procedures including dark subtraction, flat 
field division and flux calibration relative to 2MASS sources in the field. Due to the very extended 
near-infrared emission in the Serpens core, the background determination is somewhat uncertain in the 
Serpens imaging. This may cause the surface brightness to be underestimated for faint nebulosity. 
No corresponding problem was detected in the Cederblad 110 image.

In addition, mid-infrared spectrophotometry from the set of Circular Variable Filters on the Infrared 
Space Observatory Camera (ISOCAM-CVF) of Ced 110 IRS 4 was extracted from the ISO archive\footnote{In 
part based on observations with ISO, an ESA project with instruments funded  by ESA Member States 
(especially the PI countries: France, Germany, the  Netherlands and the United Kingdom.) and with the 
participation of ISAS  and NASA.}. The ISOCAM spectral image was reduced using the Cam Interactive 
Analysis (CIA) package ver. DEC01\footnote{The ISOCAM data presented in this paper were analysed using 
"CIA", a joint development by the ESA Astrophysics division and the ISOCAM Consortium. The ISOCAM 
Consortium is led by the ISOCAM PI. C. Cesarsky.}. The CVF spectrum was extracted from a 
3$\times$3\,pixel = $18\arcsec\times 18\arcsec$ aperture. The spectral resolution
is $\lambda/\Delta\lambda = R \sim 35$, but the spectrum as shown is smoothed to $R\sim 10$ to increase 
the signal-to-noise.

For CK 3, we use a 5.2-37.2\,$\mu$m spectrum obtained with the InfraRed Spectrograph (IRS) onboard the 
Spitzer Space Telescope as part of the `From Molecular Cores to Protoplanetary Disks' Legacy programme 
(c2d) \citep{Evans03}. The corresponding AOR key is 0009407232. The short-low module is used for the 
wavelength range 5.2-14.5\,$\mu$m with a resolving power of 50-100, while the short-high and long-high 
modules are used for the ranges 9.9--19.6\,$\mu$m and 18.8--37.2\,$\mu$m, respectively, with a resolving 
power of $\sim 600$. The IRS spectrum was extracted from the S9.5.0 pipeline images using the extraction 
pipeline developed by the c2d team. The details of the IRS spectrum will be discussed in more detail in 
a later paper. The short-high and long-high modules have been scaled by up to 10\% to match the short-low 
module flux. VLT-ISAAC $L$- and $M$-band spectra of CK 3 were obtained as part of a large survey of young 
low-mass stars \citep[see][and references therein]{vanDishoeck03}. The reduction procedure of the ISAAC 
spectra is described in \cite{Pontoppidan03}. The spectra cover the wavelength ranges 2.85-4.2\,$\mu$m and 
4.55--4.90\,$\mu$m in the $L$- and the $M$-band, respectively.

Near- and far-infrared photometric points were taken from the literature, specifically from 
\cite{Prusti91}, \cite{Hurt96}, \cite{Persi01} and \cite{Lehtinen01}.

\section{Radiative transfer models}
\label{RADTRANS}
We have modelled both `scenario 1' and `scenario 2b' with the Monte Carlo radiative
transfer code RADMC \citep[see][]{Dullemond04}. The code follows the photons in full 3D, but is restricted 
to an axisymmetric density structure. The same setup was also used to model the edge-on disk 
CRBR 2422.8-3423 \citep{Pontoppidan04}. Given an axisymmetric density structure, the code calculates the 
dust temperature and scattering source function at every geometric point in the model. Images at any 
wavelength can then be calculated by integrating the equation of radiative transfer along parallel lines 
of sight using the ray-tracing capabilities of the more general code RADICAL \citep{Dullemond00}. Model 
SEDs can be calculated by integrating the intensity inside synthetic `observing' apertures at each 
wavelength. This is important when comparing photometric points observed using widely differing aperture 
sizes.

A system projecting a disk shadow is modelled using a small ($\sim 100$\,AU) disk embedded in a large
dusty envelope (5\,000-10\,000\,AU) that scatters the radiation from the central disk system.
The envelope is assumed to have a radial density power law with index $\alpha_{\rm env}$. Embedded
Young Stellar Objects (YSOs) are known from theoretical models of collapsing protostars 
\citep[e.g.][]{Shu77} as well as from 1-dimensional radiative transfer model fits to millimetre wave 
emission \citep{Shirley00, Jorgensen02} to have density profiles which can be approximated by a power 
law with a negative index of 1--2. However, we allow all power law indices in the model. This can be 
justified in the case of CK 3, which probably is a much more evolved source in the process of dispersing 
the last remnants of its envelope. Infall density profiles have also been used to model scattering in 
envelopes \citep{Whitney03}. 

The geometry of the model disks is given as follows:
\begin{equation}
\rho(R,Z) = \frac{\Sigma(R)}{H_p(R)\sqrt{2\pi}}\exp\left(-\frac{Z^2}{2H_p(R)^2}\right),
\label{structEq}
\end{equation}
where $\Sigma(R)=\Sigma_{\rm disk}(R/R_{\rm disk})^{-p}$ is the surface density and 

\begin{equation}
H_p(R)/R=(H_{\rm disk}/R_{\rm disk})\times(R/R_{\rm disk})^{2/7}
\end{equation}
is the disk scale height. For $R<R_{\rm disk}$, $p\sim 1$, while for $R>R_{\rm disk}$, $p=12$. The outer 
surface density profile is arbitrarily chosen to allow a smooth transition from disk to envelope. The 
ratio $H_{\rm disk}/R_{\rm disk}$ defines the disk opening angle, 
$\Theta_{\mathrm{disk}}\equiv\tan^{-1}(H_{\rm disk}/R_{\rm disk})$. The exponent of $2/7$ in the equation for the
disk scale height is that of a passive irradiated disk in hydrostatic equilibrium with a grey dust 
opacity \citep{Chiang97}. Strictly speaking, this choice is not self-consistent, since the opacity used in the
radiative transfer is not grey. However, a more detailed discussion of the structure of flared disks is beyond the
scope of this work.
The structure of the model disk setup is described in further detail in \cite{Pontoppidan04}. 
In addition to this basic disk structure, the presence of a puffed-up inner rim is allowed by increasing 
the disk scale height within a certain radius from the central star. The physical rationale for the 
presence of a puffed-up inner rim is that the disk is irradiated at an angle of $90\degr$ at the inner disk 
edge where the dust reaches its sublimation temperature. This causes the inner rim to attain a higher 
temperature and therefore a higher pressure scale height than that of a disk irradiated at a smaller 
angle \citep{Dullemond01}. The presence of a puffed-up inner rim creates an excess near-infrared
flux which is often necessary to fit the observations both for intermediate mass Herbig Ae stars \citep{Natta01, Dominik03} and for low mass T Tauri stars \citep{Muzerolle03}. An additional effect of the inner rim, in the context of this work, is that it can produce additional shadowing. 
This may be important for CK 3 as described in \S\ref{CK3}.

A central spherical cavity of radius $R_{\rm cav}$ is carved out of the envelope. This cavity is constrained to have a radius equal to or exceeding that of the disk. No bipolar cavity is imposed, primarily because the two disk shadow candidates presented in this work do not strictly require it. Other sources may show signatures of both a bipolar cavity and a disk shadow at the same time. In the case
that the cavity has a size similar to that of the disk, it has little influence on the near-infrared morphology of the shadow. A cavity much larger than the central disk produces a `scenario 2b' disk shadow by emulating a `screen' at large distance from the central star.

A single grain size of 0.5\,$\mu$m was used, since smaller grains mostly affect scattering properties at wavelengths shorter than 1\,$\mu$m. The dust model is described in detail in \cite{Pontoppidan04}. It produces opacities similar to those for compact grains with an MRN-like size distribution by \cite{OH94} for wavelengths $>1\,\mu$m. Spherical silicate grains with inclusions of carbonaceous material are used. Ices are taken into account by adding an icy mantle consisting of mixed water, CO$_2$ and CO ices for temperatures below 90\,K using an abundance of water ice of $9\times 10^{-5}$ relative to H$_2$. No parameter concerning the ice component was allowed to be free. 
The model output is sensitive to the
exact choice of dust optical constants. However, to limit the number of free parameters in the model,
the dust properties were fixed at a single set of optical constants. It is likely that the dust in the
disk and envelope have differing optical properties due to grain growth effects and due to chemical differences \citep{Wolf03}.

Because excessive computing time prevents the calculation of a grid of models 
covering the full parameter space, the fitting parameters were changed manually and the best fit
evaluated by eye. It is therefore difficult to identify degeneracies in the model result. Yet, given a dust model, the
main fitting parameters of disk and envelope mass, luminosity, inclination and the flaring opening angle
of the disk are considered relatively robust. However, without the use of the near-infrared images, the model
would be highly degenerate. Possible degeneracies will be discussed for the specific models.

\section{CK 3}
\label{CK3}
\subsection{Observational characteristics}
\begin{figure}
   \centering
   \includegraphics[width=7.5cm]{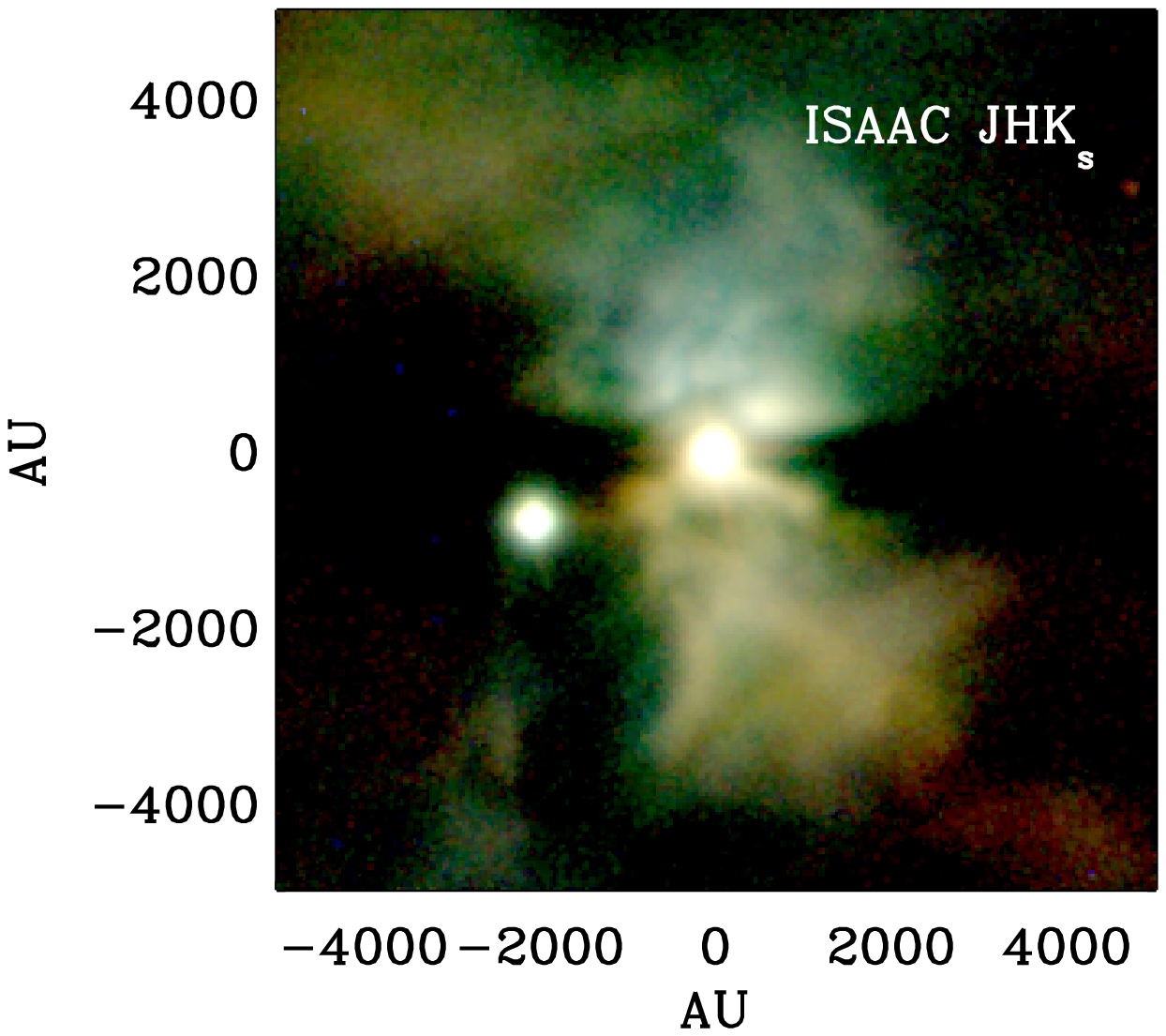}
   \includegraphics[width=7.5cm]{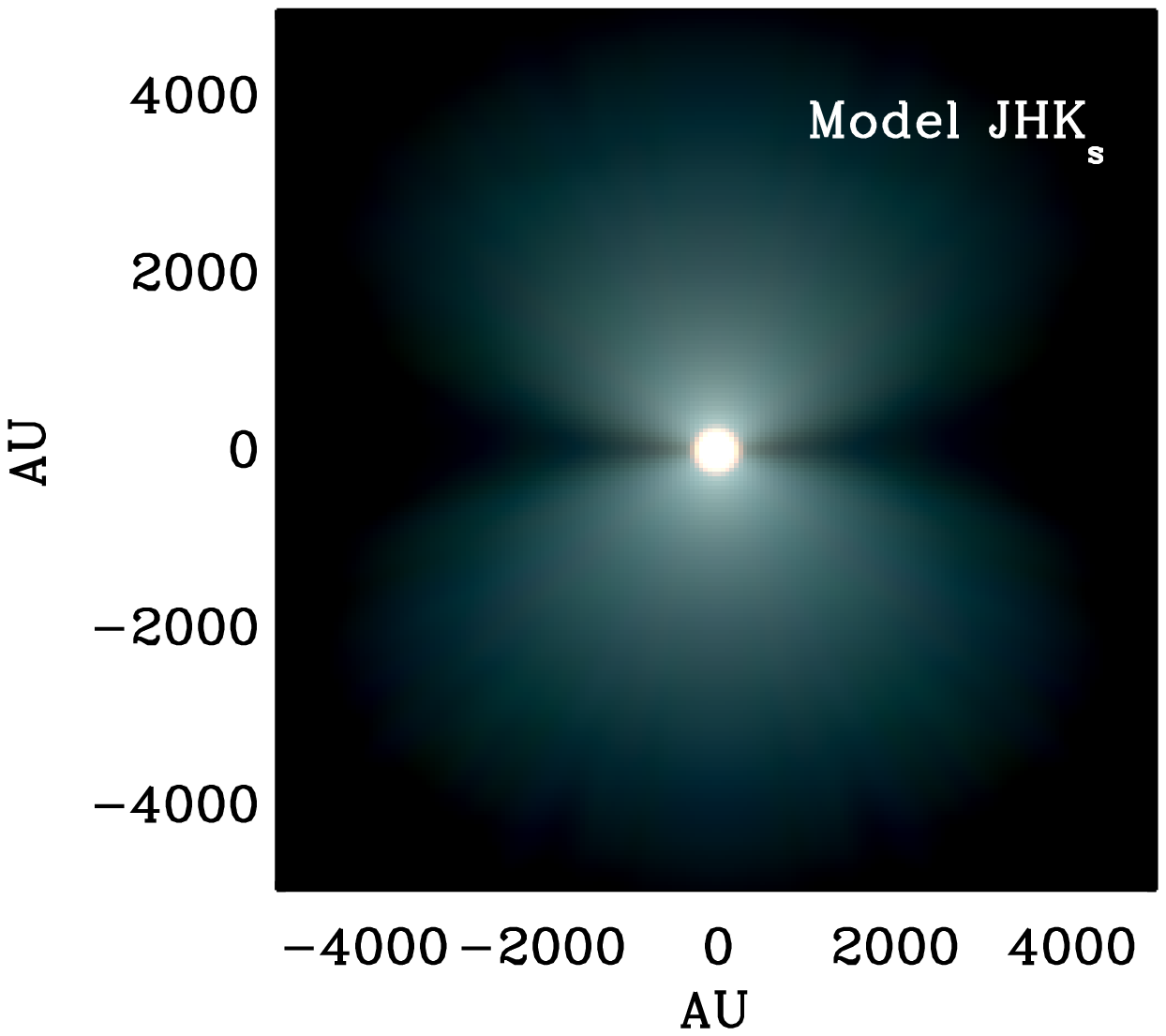}
   \includegraphics[width=7.2cm]{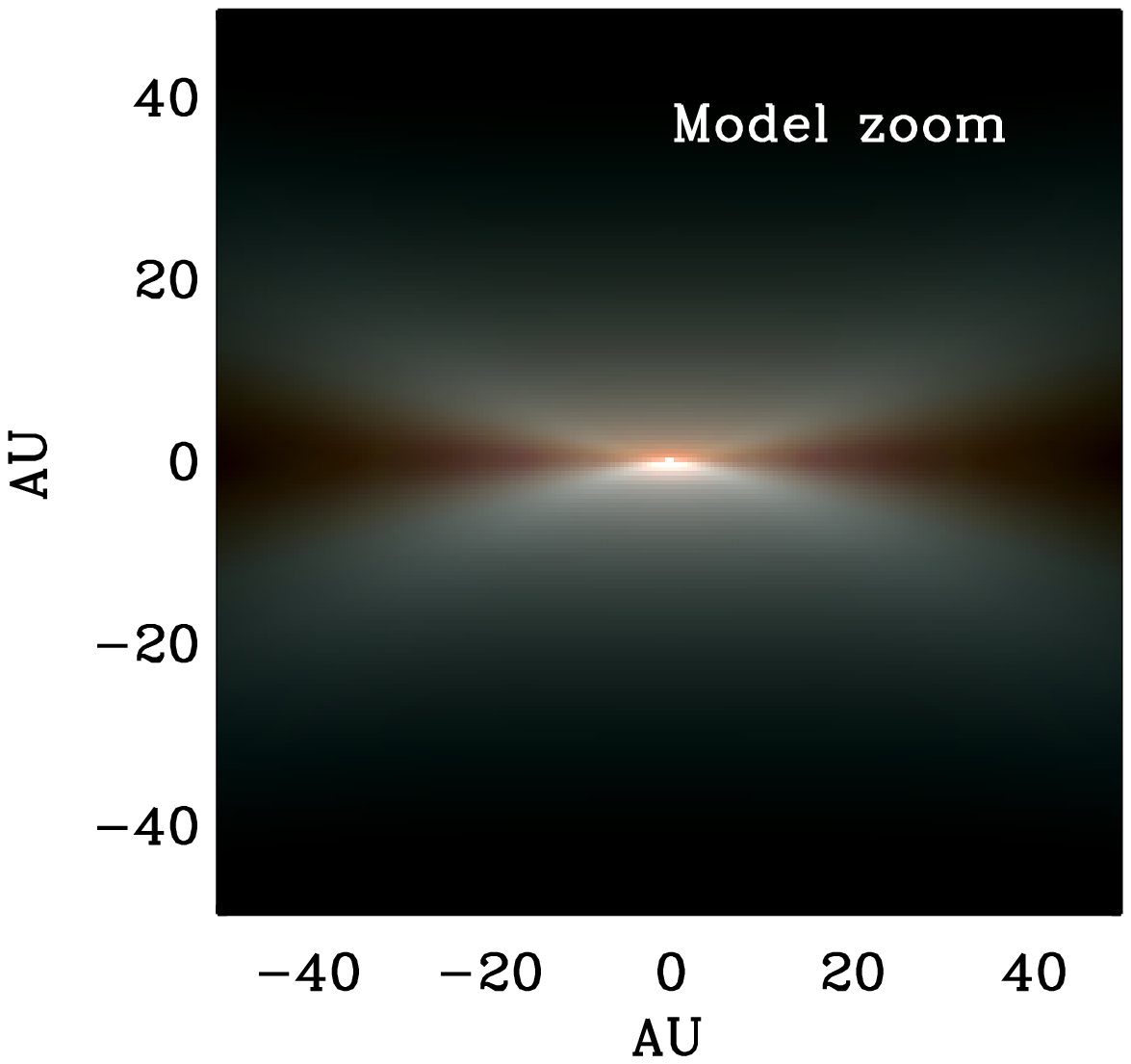}
   \caption{$JHK_s$ colour composites of the VLT-ISAAC images and the model of CK 3. All images have been constructed by assigning 
   $\log_{10}(F_{2.16\,\rm \mu m})$,$\log_{10}(F_{1.6\,\rm \mu m})$ and $\log_{10}(F_{1.25\,\rm \mu m})$ to the red, green and blue colour channels, respectively.  {\it Left panel:} Observed ISAAC $JHK_s$ colour composite. The observed image has been rotated by 47\degr east of north. {\it Right panel:} The best-fitting model of CK 3. The model has been convolved with a Gaussian with a $FWHM$ of $0.7\arcsec$ to match the observed image quality. {\it Bottom panel:} Zoomed-in view of the physical disk of the model on a 100\,AU scale. The images are aligned
   with north pointing up and east to the left.}
  \label{images_EC82}   
\end{figure}

The Serpens reflection nebula is characterised by a bright central compact source (CK 3) surrounded by a reflection nebulosity several arcminutes in extent. The distance to Serpens was measured by \cite{Straizys96} to 260\,pc and to 220--270\,pc by Knude et al., in prep. All distance dependent 
quantities for CK 3 are scaled to 250\,pc. Near-infrared imaging polarimetry has shown CK 3 to be the illuminating source of most of the Serpens reflection nebulosity \citep{Sogawa97,Huard97}. The nebulosity  
is entirely bisected by a dark lane centered on CK 3. The angular extent of the lane is at least $35\arcsec$, which corresponds to more than 9000\,AU. In the ISAAC images, the central source is extended in the $H$ and $K_s$ bands 
with deconvolved $FWHM$ of $0\farcs5=125$\,AU and $0\farcs8=200$\,AU, respectively. This may mean than the disk is
resolved in the images, providing a lower limit to the physical disk size.
The $JHK_s$ colour composite image of CK 3 and the surrounding nebulosity is shown in Fig. \ref{images_EC82}. It is seen that the dark lane is shaped as two symmetric wedges with completely straight edges. The border between the dark lane and the nebulosity is unresolved at a spatial resolution of $\sim 0\farcs6$.
\cite{Sogawa97} observed significant near-infrared variability of the nebulosity illuminated by CK 3 from 1991 to 1992 of up to 1\,mag\,arcsec$^{-2}$ in the $H$ band. The strongest variation is observed along the dark lane.
This is of particular interest to the disk shadow scenario, since the dynamical timescale of the inner disk
suggested to be casting a shadow is on the order of days to years, while nebulosity on scales of 5\,000--10\,000\,AU 
should have significantly longer dynamical time scales. The implications of this for the model are further discussed in the next section.

We suggest that the dark band in the Serpens reflection nebula is a shadow of a much smaller disk surrounding CK 3. A real edge-on disk will tend to be shaped as a box with reflection nebulosity along the minor axis of the disk rather than a wedge. An example of such a disk is the Butterfly Nebula \citep{Wolf03}. The sharpness of the edge of the dark band also resembles a shadow created by photons travelling in a straight line from the central star. A real disk may be
expected to show some asymmetric structure on 10\,000\,AU scales. We propose that CK 3 and the associated Serpens reflection nebula represent an example of a `scenario' 1 disk shadow, i.e. a disk shadow cast into a spherically symmetric, volume-filling envelope.

\begin{figure}
   \centering
   \includegraphics[width=8cm]{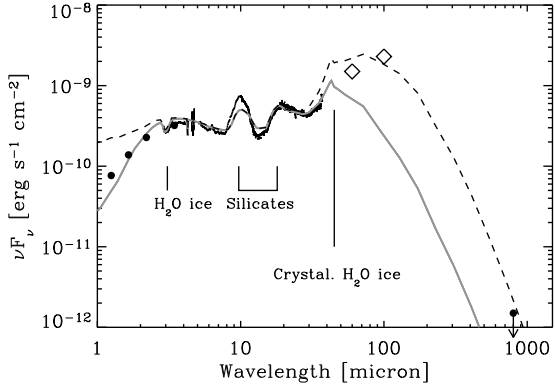}
   \caption{The model fit to the observed SED of CK 3. The solid curve is the SED measured through an aperture
   of 5\arcsec, while the dashed curve is the SED measured through a 60\arcsec aperture to match the IRAS photometric points
   (indicated by diamonds).}
   \label{sed_ec82}
\end{figure}

An important question to explore is if the SED of the source
as well as the near-infrared morphology can be fitted by a disk shadow model. 
The observed SED of CK 3 is shown in Fig. \ref{sed_ec82} including the
3--5\,$\mu$m VLT-ISAAC spectra and the 5.2--37.2\,$\mu$m Spitzer-IRS spectrum. The overall shape of the SED is that of a class II or flat spectrum source, typical of a young star surrounded by a disk. 
CK 3 is dominated in the mid-infrared by very bright emission features from silicates at 10 and 18\,$\mu$m. Additionally,  PAH emission features are seen at 3.3 and 6.2\,$\mu$m and broad ($\sim 80\,\rm km\,s^{-1}$) ro-vibrational emission lines from warm, gaseous CO are observed around 4.7\,$\mu$m, while shallow absorption bands from water and CO ices are evident at 3.08 and 4.67\,$\mu$m. The shapes of the water and CO ice bands indicate that the ice is cold ($\lesssim 30\,$K). The far-infrared photometric points are taken from IRAS measurements by \cite{Hurt96}. \cite{Casali93} only found an upper limit at 800\,$\mu$m of 0.4\,Jy. It should be cautioned that the Serpens core is a very confused region
in the 60\,$\mu$m IRAS beam of $1\arcmin$. The IRAS fluxes have been colour corrected to match
the observed $\sim 30\,$K spectral slope with factors of 1.1 and 1.02 at 60 and 100\,$\mu$m, respectively \citep[see the IRAS explanatory supplement,][]{IRAS}.

 The wealth of emission features in the spectrum presents a challenge to a disk shadow model since strong silicate emission is surprising from a highly inclined disk, given the expected large optical depth along the disk mid-plane. Assuming the silicate emission is coming from the inner 10\,AU of the disk, the features can be explained only if the disk is optically thin to 10\,$\mu$m photons along the line of sight. This requires the disk to be very tenuous, at least in small dust grains. Significant mass may
be hidden in large, millimetre-sized dust grains, which are not probed in the mid-infrared. However, to
create the near-infrared shadow, the disk must be optically thick to 1-2\,$\mu$m photons along the same line of sight. This puts severe constraints on the model disk structure, and in particular on the small grain dust mass.

\subsection{Model of CK 3}

\begin{figure*}
   \centering
   \includegraphics[width=6.5cm]{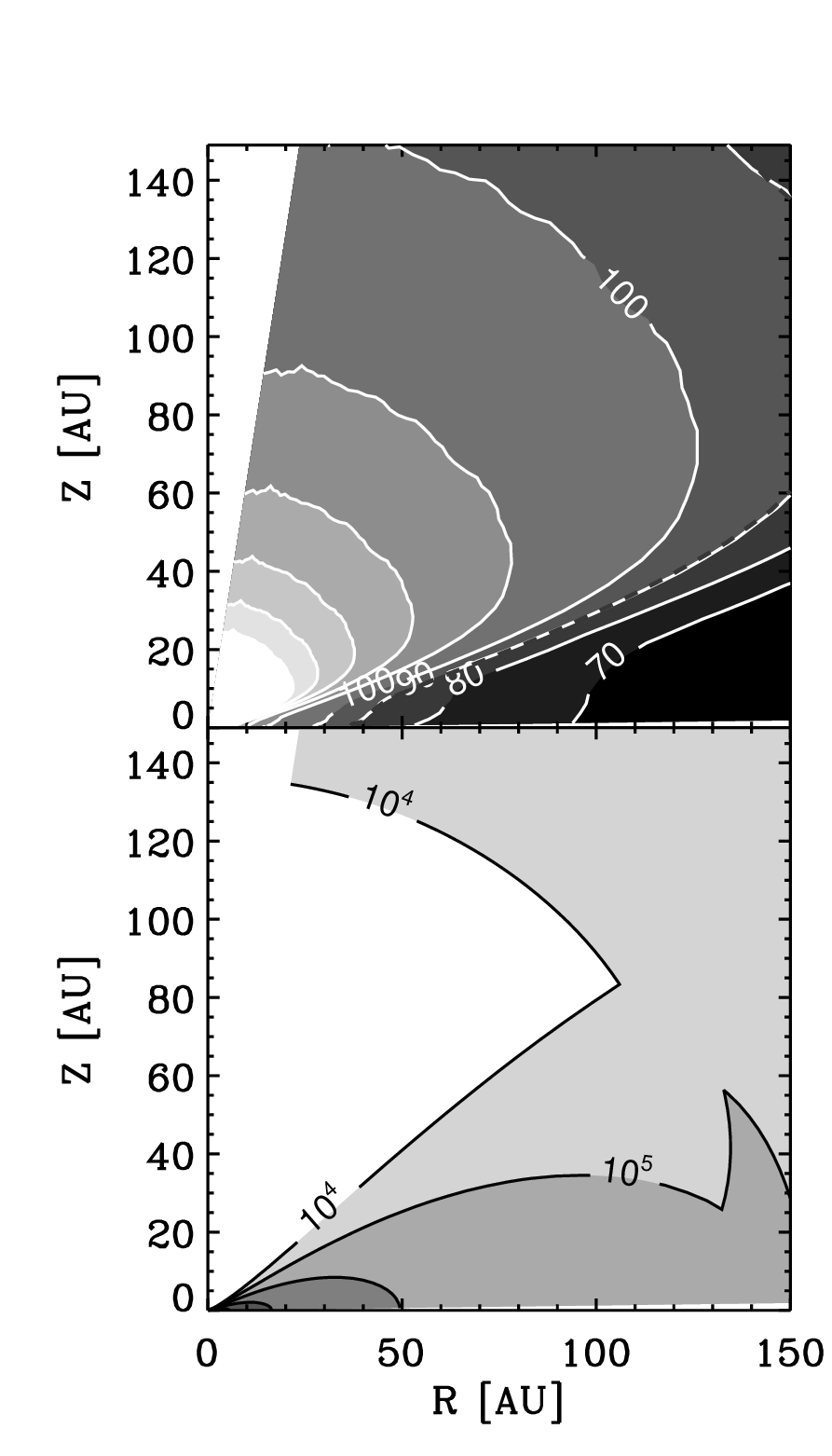}
   \includegraphics[width=6.5cm]{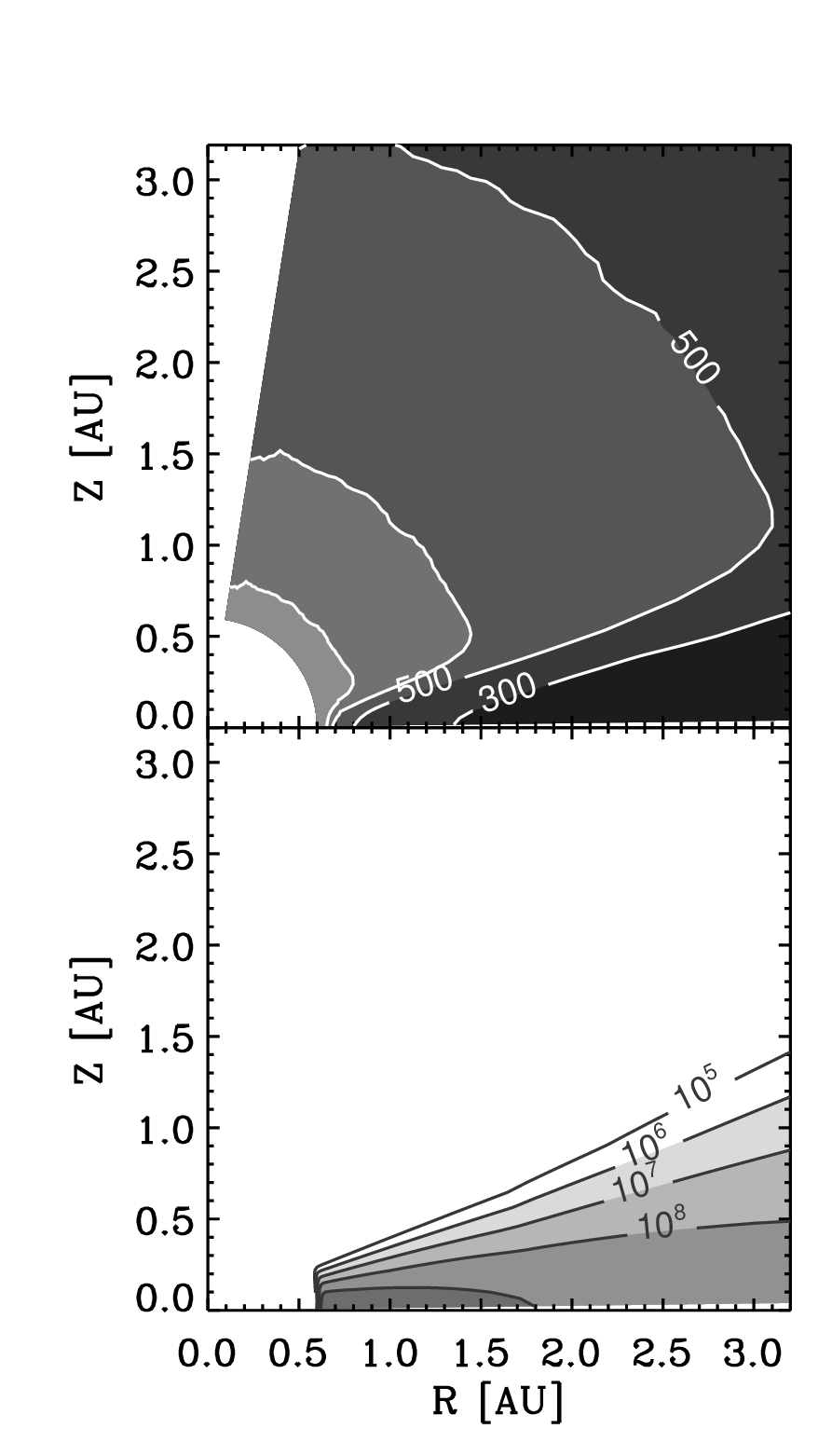}
   \caption{Temperature and density structure of the CK 3 model. The top panels show the temperature in units of Kelvin
   and the lower panels show the density in units of cm$^{-3}$. The left panels show the entire disk, while the
   right panels show the inner parts of the disk.}
  \label{struct_EC82}   
\end{figure*}

The best-fitting parameters of the CK 3 model are given in Table \ref{ModelPars} and the model temperature and density structure
are shown in Fig. \ref{struct_EC82}. The resulting model $JHK_s$ colour composite is compared to the observed colour composite in 
Fig. \ref{images_EC82}. 

\begin{table}
\centering
\caption{Model parameters for the two disk shadow candidates}
\begin{tabular}{lll}
\hline 
\hline 
 & CK 3/EC 82 & Ced 110 IRS4  \\
\hline
$M_{\rm disk}  [M_{\odot}]$&$7.5\times 10^{-6}$&$3.5\times 10^{-5}$\\
$M_{\rm env}  [M_{\odot}]$&0.20&0.12\\
$R_{\rm cav}$ [AU]&140&530\\
$L_{\rm bol} [L_{\odot}]$&9.9&0.4\\
$R_{\rm env}$ [AU]&6000&4000\\
Incl. angle&77$\degr$&72$\degr$\\
$T_{\rm star}$ [K]&6500&4800\\
$M_{\rm star} [M_{\odot}]$&2.1&0.5\\
$R_{\rm disk}$ [AU]&140&35\\
$\alpha_{\rm env}$&-0.05&-1.5\\
$H_{\rm disk}/R_{\rm disk}$&0.3&0.28\\
$H_{\rm rim}/R_{\rm rim}$&0.07&0.29\\
$p$&0.8&1.0\\
\hline
\end{tabular}
\label{ModelPars}
\end{table}

\begin{figure*}
   \centering
   \includegraphics[height=132mm,angle=90]{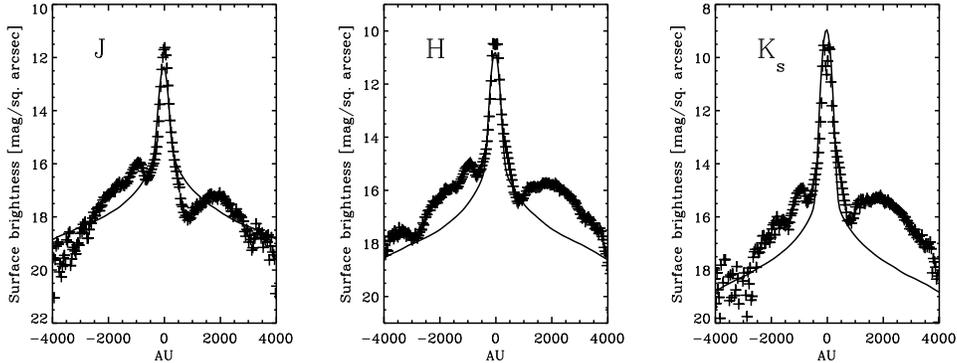}
   \caption{Cross section along the minor axis of CK 3 in the different near-infrared wavebands. The curve shows the model, while the (+) symbols show the observed ISAAC surface brightness. Clumpy envelope structure not included in the model
   gives excess scattering at radii of $\sim 2000$\,AU. }
   \label{cross_EC82}
\end{figure*}

\begin{figure*}
\centering
   \includegraphics[width=6.5cm]{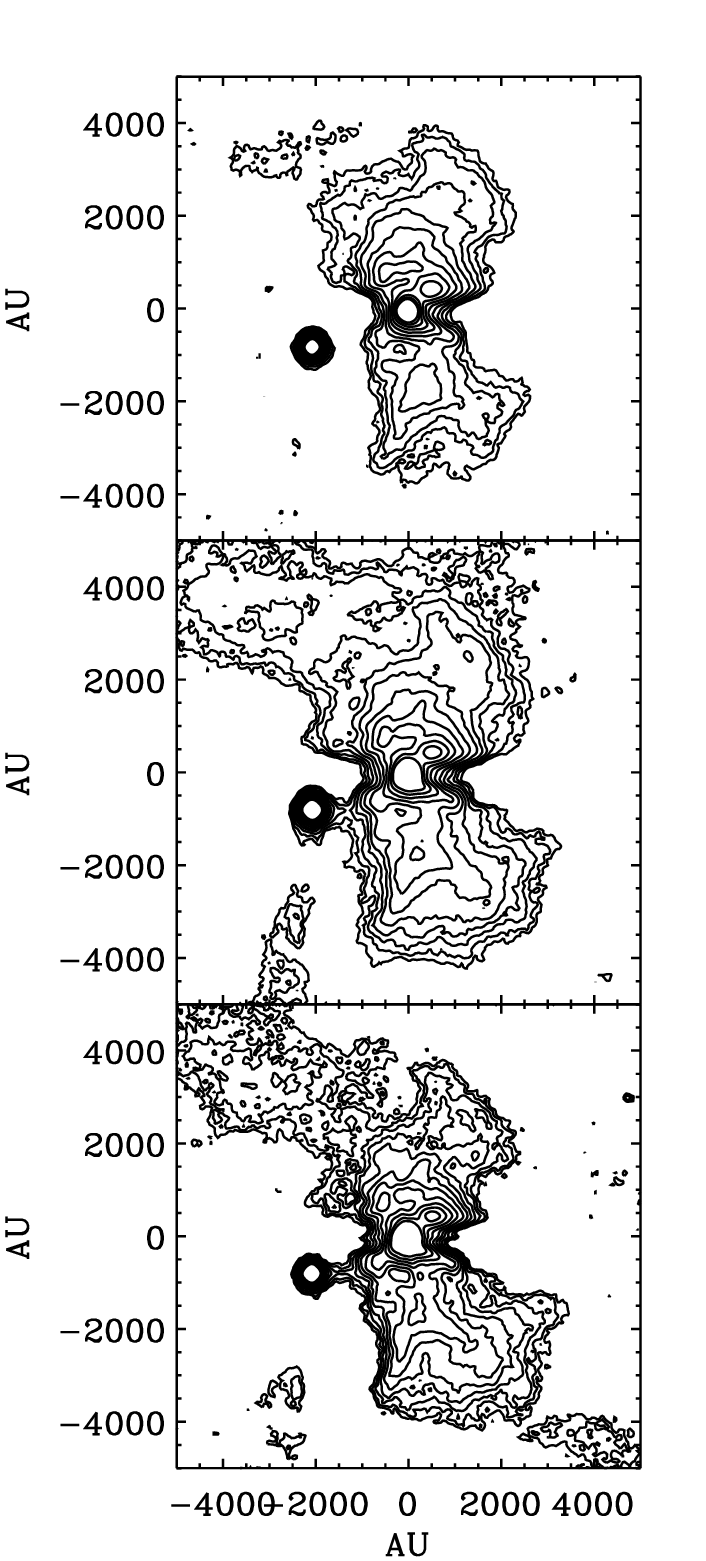}
   \includegraphics[width=6.5cm]{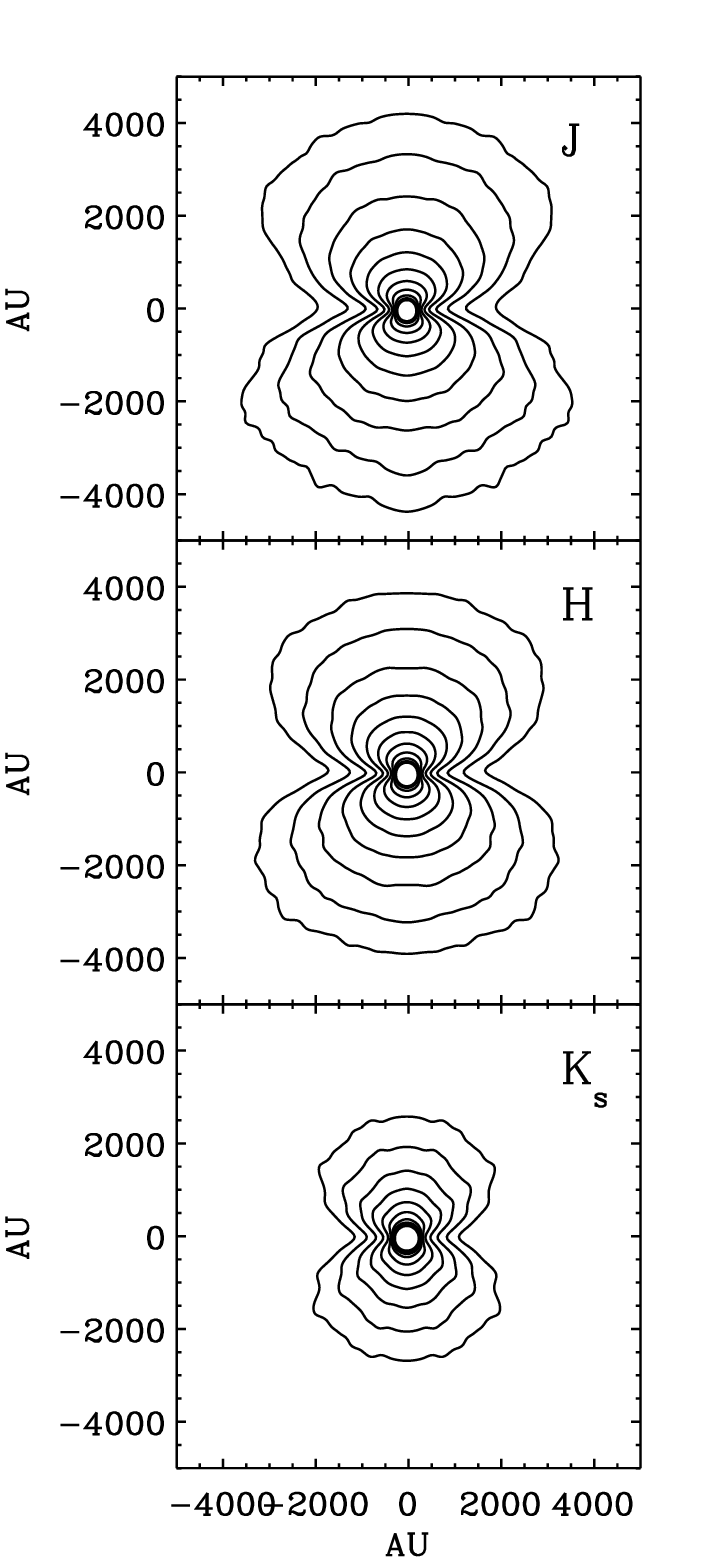}
 \caption{Contours of the observed (left) and modelled (right) $JHK_s$ band images. Model and observation are plotted
 on the same intensity scale. The contours are spaced logarithmically in steps of 0.2 dex with the lowest contour at $\rm 1.7\times 10^{-17}$\,erg\,s$^{-1}$\,cm$^{-2}$\,Hz$^{-1}$\,sterad$^{-1}$. The source to the south-east of CK 3 is the unrelated star EC 86 which is probably located in the foreground of the
 envelope material.}
 \label{EC82_contour}
\end{figure*}

Following the hypothesis that CK 3 is a `scenario 1' disk shadow,
the object is modelled as a disk surrounded by an envelope extending inwards to the outer disk edge. The sudden increase in flux at wavelengths longer than 40\,$\mu$m as evidenced by the SED (Fig. \ref{sed_ec82}) indicates that
the amount of warm dust in the envelope is small. This in turn requires a very flat density profile of the envelope. A nearly constant envelope density provides the best fit to the far-infrared SED and the extended emission seen in the near-infrared images (Fig. \ref{images_EC82}). The model fit to the near-infrared nebulosity
is shown in quantitative terms in Fig. \ref{cross_EC82}, where a cross section along the minor axis of the system is shown for each near-infrared band and in Fig. \ref{EC82_contour}, which compares contours of modelled and observed images. It is seen that the reflection nebulosity contains a significant amount of clumpy structure at radii larger than $\sim 1000$\,AU that the model does not take into account. However,  the power law envelope provides
a reasonable fit to the underlying structure. The clumpy envelope structure is likely to
affect the far-infrared SED by increasing the fraction of cold dust. Such structure cannot be accurately modelled by the current axisymmetric model setup, but may explain the relatively poor fit to the IRAS photometric points. 
The best-fitting envelope has a column density to the disk of $N({\rm H}_2)=4\times 10^{21}\,$cm$^{-2}$, corresponding to an extinction of $A_J=1.4\,$mag. This is consistent with the observed column density in the 3.08\,$\mu$m water ice band \citep{Pontoppidan04b}, assuming an ice abundance of $9\times 10^{-5}$ relative to H$_2$. 

The disk itself is modelled with $R_{\rm disk}=140$\,AU and an outer scale height of $H_{\rm disk}/R_{\rm disk}=0.3$.
The physical property probed by the observed disk shadow opening angle is the angle relative to the disk plane at which the disk becomes optically thick to near-infrared photons. This depends on the vertical structure
of the dust in the disk, the disk opening angle, $\Theta_{\mathrm{disk}}\equiv \tan^{-1}(H_{\rm disk}/R_{\rm disk})$, as well as the surface density profile $\Sigma(R)$ of the disk.
The observed wedge-angle also depends on the disk inclination, getting smaller at lower inclinations. The edges of the shadow are only straight for high
inclinations of more than $\sim 75\degr$. This effectively constrains the inclination to $>75\degr$. At very high inclinations the near-infrared colours  of the central source become too red, such that the best fit of the inclination is
77\degr. The wedge shadow disappears completely for inclinations less than $90\degr-\Theta_{\mathrm{disk}}$. This means that a `scenario 1' shadow is observed only when the line of sight to the central star passes through the disk, either the puffed-up
inner rim or the outer flared part of the disk. If the disk has a high optical depth then the star is fully obscured. For intermediate optical depths, which are still high enough to cast a shadow but low enough to allow some starlight to penetrate the disk, the star may still be seen through the disk. In the case of CK 3, the SED of the source, if viewed in isolation, can be equally well fitted with a face-on disk and only the presence of the disk shadow fixes the inclination. 

Therefore, in order to produce the bright silicate emission features (see Fig. \ref{sed_ec82}) and at the same time maintain 
that the line of sight to the central star passes through the disk as given by the presence of a disk shadow, the extinction 
toward the inner parts of the disk must be very low. At the same time, the 1-2\,$\mu$m optical depth must be higher than unity 
in order to cast the shadow. The assumed density structure in the disk (Eq. \ref{structEq}) then produces a very low total disk 
mass. Note, however, that the mass refers only to the mass in small dust grains. Effects such as dust growth or settling may 
produce similar effects for a more massive disk if most of the dust grains have grown to sizes $\gg 10\,\mu$m. Large dust grains 
tend to settle into the disk mid-plane \citep{Dubrulle95}. Such `pebble-sized' grains are not probed by the infrared observations. 
The strength of the silicate emission features indicates that grains much larger than $\sim 1\,\mu$m do not contribute to the 
mid-infrared opacity of the warm dust. A dust grain size distribution dominated by intermediate-sized (1--10\,$\mu$m) 
grains suppress solid state features \citep{Boekel03}. This means that if the  actual disk has a dust mass much larger than 
that of the model, the average grain size distribution in the disk must be highly bimodal, i.e. a deficiency of 
intermediate-sized 1--10\,$\mu$m grains must be present. An additional population of very large ($>$cm) grains, perhaps even 
containing most of the total dust mass, may be assumed to reside in the midplane without having any effect on the resulting 
model images or SEDs. For lack of observational constraints, our model disks contain only small (0.5 $\mu$m) grains. The best 
fit is obtained for a disk mass of $7.5\times 10^{-6}\,\rm M_{\odot}$, assuming a gas-to-dust ratio of 100 and 0.5\,$\mu$m 
grains. This creates bright silicate emission features as well as a disk shadow. A more massive disk, with a mid-plane that 
is optically thick at all mid-infrared wavelengths, is able to produce at most 1/3 of the observed silicate line-to-continuum 
ratio, even when observed face-on. The unusual strength of the emission features is therefore strong evidence for a very 
tenuous, almost optically thin disk. The disk mid-plane between 50-150\,AU has gas densities of $10^5-10^6\,$cm$^{-3}$ 
(Fig. \ref{struct_EC82}). Again, if dust settling has taken place, the actual gas density in the disk may be higher. 
Interferometric observations of molecular gas-phase lines may answer this question. The strength of the silicate emission 
features as well as the depth of the shadow are dependent on the disk surface density profile. It is found that a surface 
density profile with $p=0.8$ produces the strongest shadow and emission features. As a side effect of the low optical depth 
of the disk, the model spectrum shows strong emission from crystalline water at 45\,$\mu$m, a band which unfortunately lies 
outside the coverage of Spitzer-IRS. 

The model fit has some weaknesses that are important to address. In the model, the shadow is not completely dark because of the 
relatively low optical depth through the disk required by the silicate emission features. Alternatively, at high envelope 
densities or if large grains are present, the shadow can be filled out by multiple scattering, although this is not the case 
for the CK 3 model. The puffed-up inner rim may play in important role for the
shadow of CK 3. Because the outer disk is so tenuous, the rim can provide a significant part of the column density producing 
the shadow. The shape and strength of the shadow is therefore very sensitive to the scale height of the puffed-up inner rim. 
In this context, it is intriguing that \cite{Sogawa97} found the Serpens reflection nebulosity to be variable on timescales 
of up to a year. If the scale-height of the puffed-up inner rim is variable, the opening angle and strength of the shadow 
may change dramatically accompanied by a strong variability in the 1-5\,$\mu$m wavelength region of the SED. 
The spectral data may therefore not correspond completely to the ISAAC imaging, since the observations have been obtained 
at different epochs. The inner rim in the model has a temperature of 900\,K in order to fit the shape of the 2-6\,$\mu$m 
spectrum. As seen in Fig. \ref{struct_EC82}, this corresponds to the presence of a gap in the disk with a radius of 0.6\,AU. 
The location of the inner rim at this distance from the central star is consistent with a time scale of months for 
variations of the shadow produced by the inner parts of the disk. While only a fraction of the model shadow is produced
by dust within 1\,AU in the present model, an increase of the scale-height of the inner rim of $\sim 50\%$ is able
to cause easily observable differences in the surface brightness of the reflection nebula of several magnitudes in the shadowed 
region. However, a change in the scale height of the inner rim also causes considerable changes in the strength of the
silicate emission features. 
Other disk shadows may be variable on time scales ranging from days to years, depending on the size of the central dust gap.
It is noted that the observed width of the CO fundamental ro-vibrational emission lines in the ISAAC $M$-band spectrum 
of $\sim 80\,$km\,s$^{-1}$ is consistent with gas at $\sim 1\,$AU in an edge-on disk in Keplerian rotation.

\section{Ced 110 IRS 4}
\label{CED}
\subsection{Observational characteristics}
Ced 110 IRS 4 is a class I YSO embedded in the Cederblad 110 cloud of the Cha\-maeleon molecular cloud at a distance of 
150\,pc \citep{Knude98}. The region contains a large scale outflow \citep{Mattila89}. The source of the outflow has not 
been unambiguously identified, although both Ced 110 IRS 4 as well as the embedded protostar Cha MMS1 have been proposed 
\citep{Reipurth96}. Recently, \cite{Lehtinen03} determined that Ced 110 IRS 4 is a strong continuum radio source in the 
centimetre wavelength region and noted that this is evidence that IRS 4 is the source of the outflow. However, the outflow 
mapped in $^{12}$CO by \cite{Prusti91} is centered closer to the position of Cha MMS1 than that of IRS 4.

\begin{figure}
   \centering
   \includegraphics[width=7.0cm]{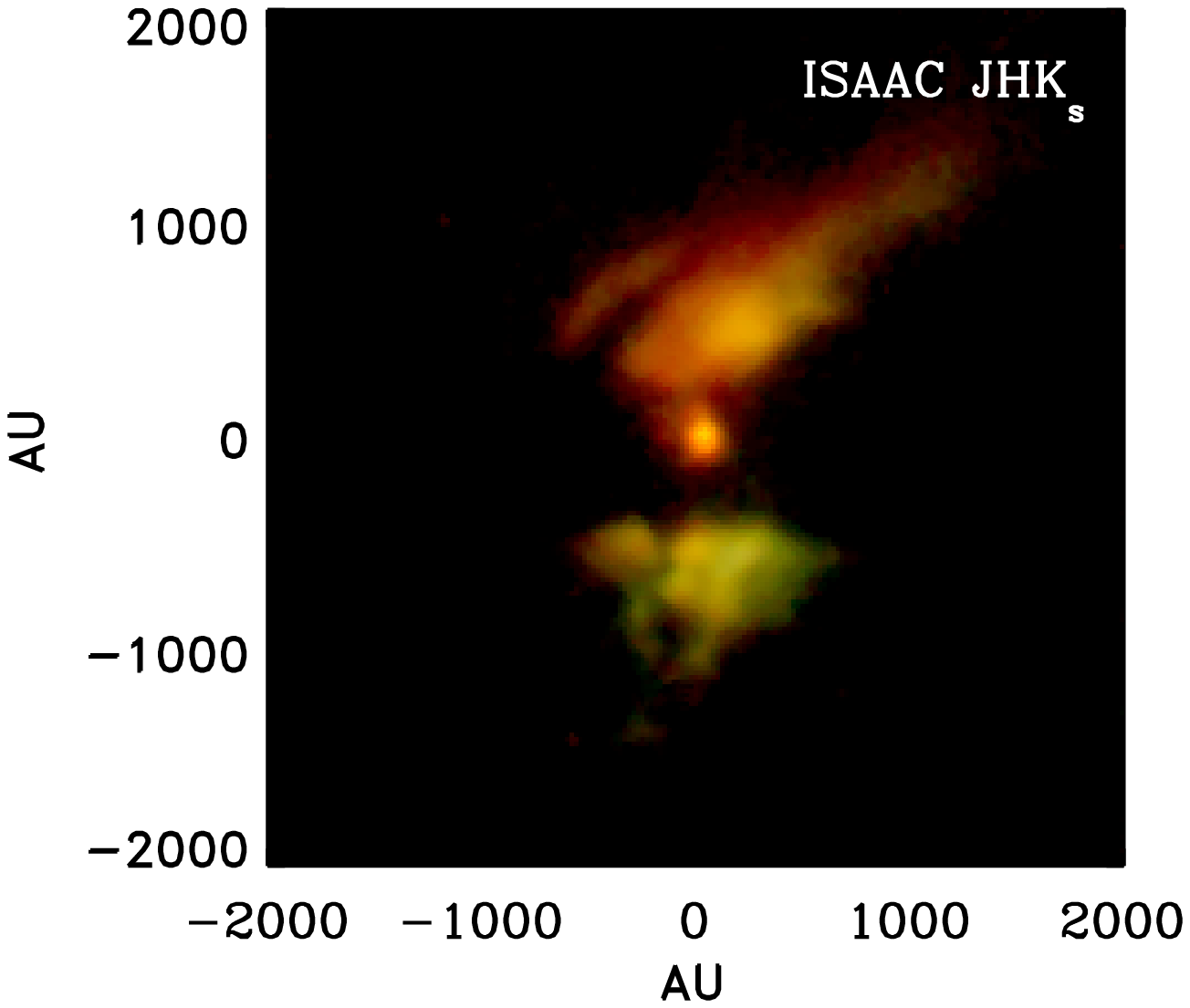}
   \includegraphics[width=7.0cm]{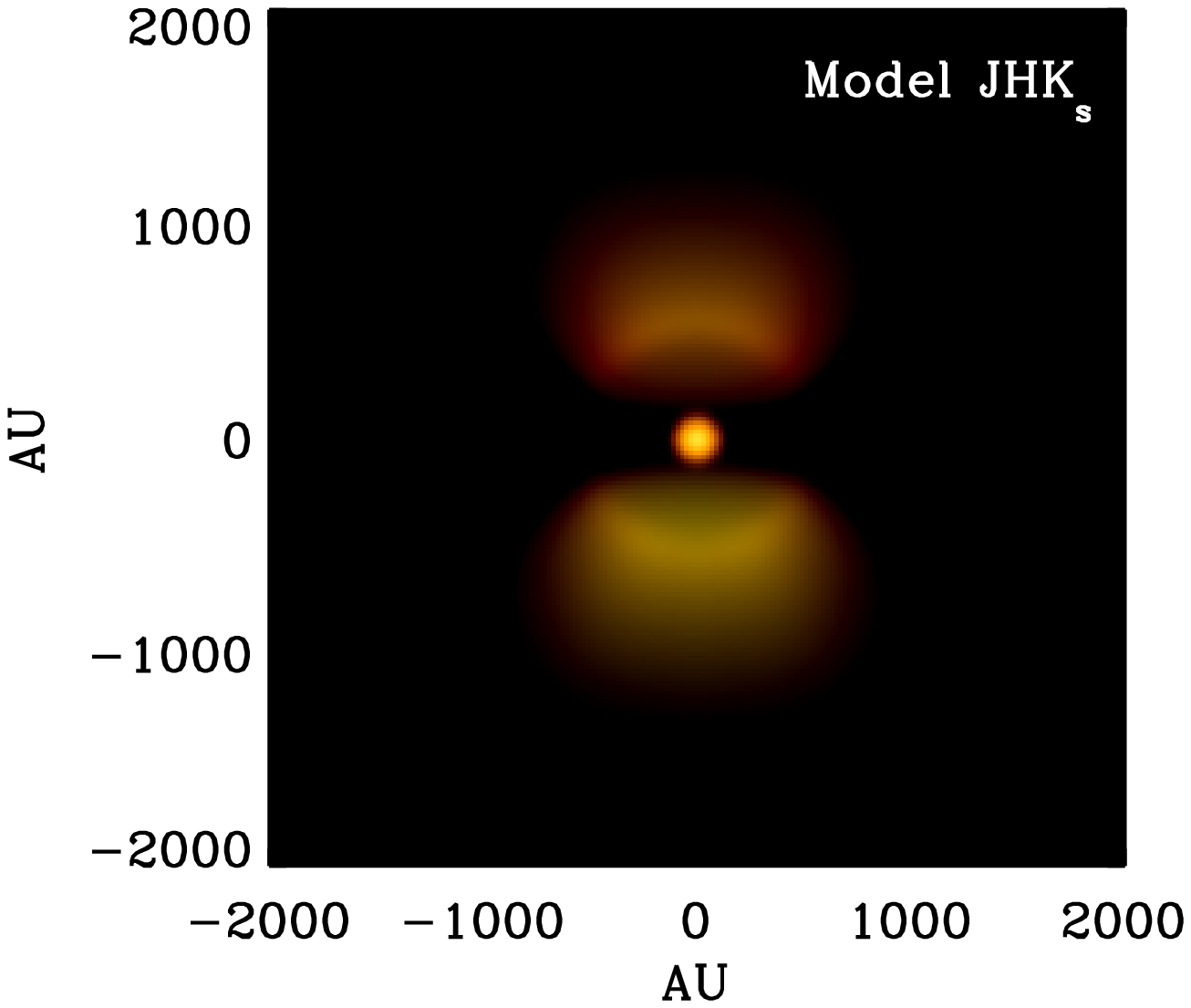}
   \includegraphics[width=6.7cm]{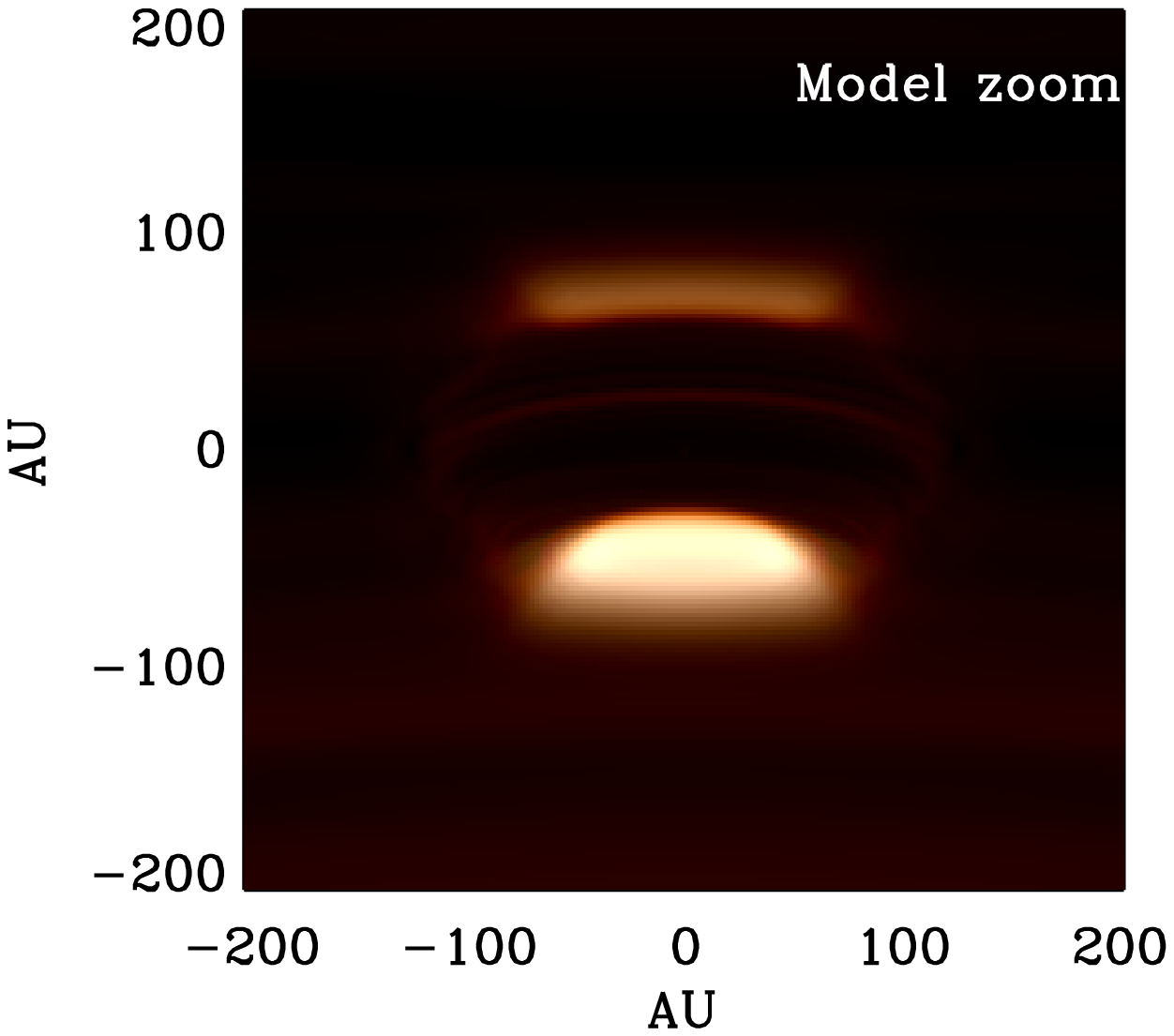}
   \caption{$JHK_s$ colour composites of the VLT-ISAAC images and the model Ced 110 IRS 4 constructed as in Fig. \ref{images_EC82}. 
{\it Top panel:} Observed ISAAC $JHK_s$ image. {\it Middle panel:} The model $JHK_s$ image of Ced 110 IRS 4 convolved with a 
Gaussian with a $FWHM$ of $0.7\arcsec$ to match the observed image quality. {\it Bottom panel:} A zoom-in view of the physical disk 
of the model on a 100\,AU scale. The images are aligned
   with north pointing up and east to the left.}
   \label{images_ced}
\end{figure}

The VLT-ISAAC near-infrared images are shown in Fig. \ref{images_ced}. The images show a bipolar nebulosity with
lobes completely separated by an almost rectangular dark band. A compact source is seen near the center of the dark lane. 
From the near-infrared colours, it is seen that the central compact source as well as the nebulosity are strongly 
reddened \citep{Zinnecker99}. The thickness of the lane in the north-south direction ($\sim 4.5\arcsec=675$\,AU) 
relative to the width of lane ($\sim1000\,$AU) seems large for a physical disk. The edges of the lane are sharp and 
unresolved at the $\sim 0.7\arcsec$ seeing of the ISAAC images. The outer edges of the reflection nebula also show a 
very steep decline, in particular the southern lobe. The northern lobe of the bipolar nebula is significantly more 
reddened than the southern lobe. The central source is observed to be extended measured relative to other stars in 
the field in the ISAAC $H$- and the $K_s$-bands with deconvolved $FWHM$ of 0.5\arcsec=75\,AU and 0.4\arcsec=60\,AU, 
respectively. The physical size of the reflection nebulosity of Ced 110 IRS 4 is about half that of the CK 3 nebula. 
The $K_s$-band surface brightness is similar, but the central source is clearly much more extincted. An obvious difference 
is the morphologies of the two dark bands; the dark band of Ced 110 IRS 4 is clearly not wedge-shaped as the band of CK 3 is.

\begin{figure}
   \centering
   \includegraphics[width=8cm]{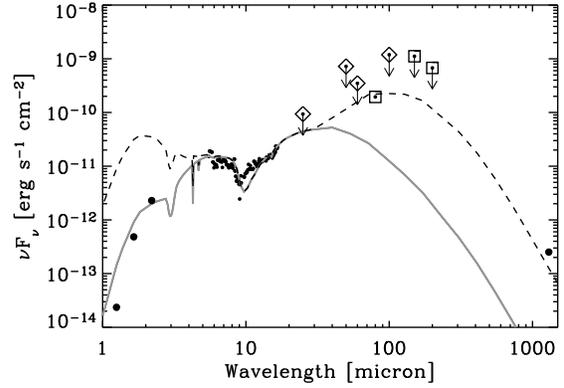}
   \caption{The model fit to the observed SED of Ced 110 IRS 4. The solid curve is the model SED through
   a 2\arcsec aperture, while the dashed curve is the model SED through a 50\arcsec aperture. Upper limits
   indicate detections blended with nearby sources. These include 25, 60 and 100\,$\mu$m IRAS points
   as well as 150 and 200\,$\mu$m ISOPHOT points. The photometric points are taken from \cite{Prusti91} (IRAS points indicated by
   diamonds) and \cite{Lehtinen01} (ISOPHOT points indicated by squares). }
   \label{sed_ced}
\end{figure}

The SED of Ced 110 IRS 4 is shown in Fig. \ref{sed_ced}. 
It is composed of near-infrared photometry of the central source (excluding the nebulosity), an archival 
ISOCAM-CVF 5-16\,$\mu$m spectrum, ISOPHOT photometry at 80-200\,$\mu$m as well as IRAS photometry and a 1.3\,mm 
point from \cite{Prusti91}. The source is bright at far-infrared to millimetre wavelengths as measured by ISOPHOT 
\citep{Lehtinen01}, indicative of a significant amount of remnant envelope material.
The far-infrared points are very scattered and do not produce a very well-defined SED. This is probably due to source 
confusion as noted by \cite{Lehtinen01}. Specifically, Lehtinen et al. suggest that a previously unidentified 
far-infrared source designated IRS 11 is located about $30\arcsec$ to the south-east of Ced 110 IRS 4. Only the 
80\,$\mu$m point has the contribution from IRS 11 removed. Therefore, all photometric points which are blended are 
used as upper limits and the 80\,$\mu$m and the 1.3\,mm points are used to constrain the model. The ISOCAM spectrum 
shows a fairly shallow silicate 9.7\,$\mu$m absorption band. The low signal-to-noise of the
spectrum prevents any other features to be identified.

\subsection{Model of Ced 110 IRS 4}
A disk shadow model of Ced 110 IRS 4 must reproduce the broad dark band and the central compact source seen in the near-infrared 
image (Fig. \ref{images_ced}). This morphology is not consistent with a volume-filling dust-distribution, but is most easily explained 
with a spherical distribution of dust with a central spherical cavity, corresponding to `scenario 2b' in 
Fig. \ref{screen_cartoon}.

We therefore suggest that the morphology of Ced 110 IRS 4 can be explained as a shadow of a small, inclined disk projected on the 
remnant envelope by introducing an empty {\it spherical} cavity with a radius of 530\,AU. Note that this is not a {\it bipolar} 
cavity, often suggested to be required to produce a bipolar reflection nebulosity. 
Within the radius of the spherical cavity centered on the star-disk system, the shadow will be a dark band with parallel edges. 
At radii larger than the cavity, the shadow will continue the wedge-like shape as in the case of a `scenario 1' source. 
In essence, the scattering material will appear to an observer as a bipolar set of truncated cones, but without the 
need to invoke a bipolar cavity in which the light can be scattered. In some cases, it may be difficult to distinguish a 
disk shadow from a bipolar cavity observationally. However, the central compact near-infrared source is a clear signature 
of a projected disk shadow. 

\begin{figure*}
   \centering
   \includegraphics[width=6.5cm]{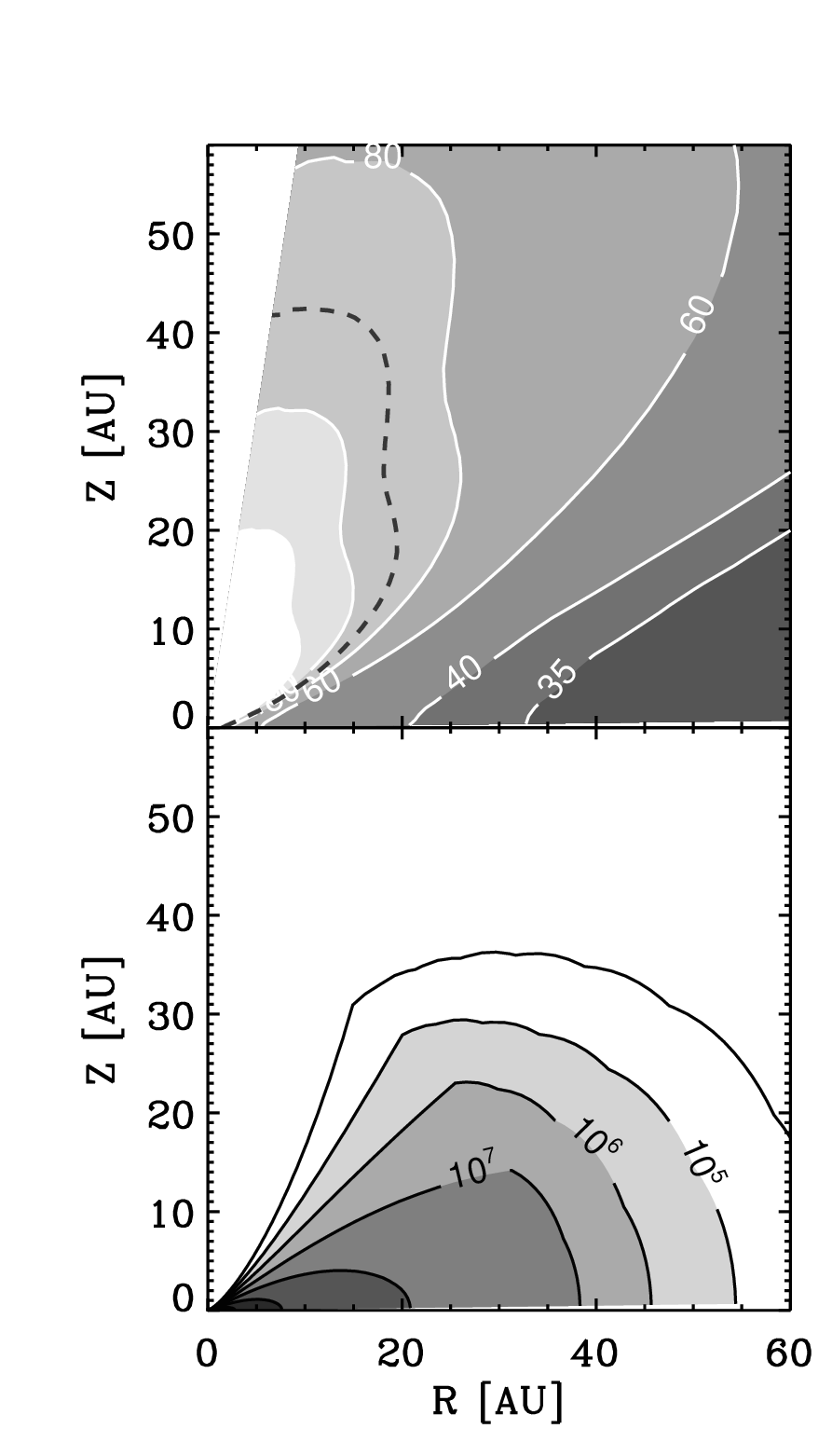}
   \includegraphics[width=6.5cm]{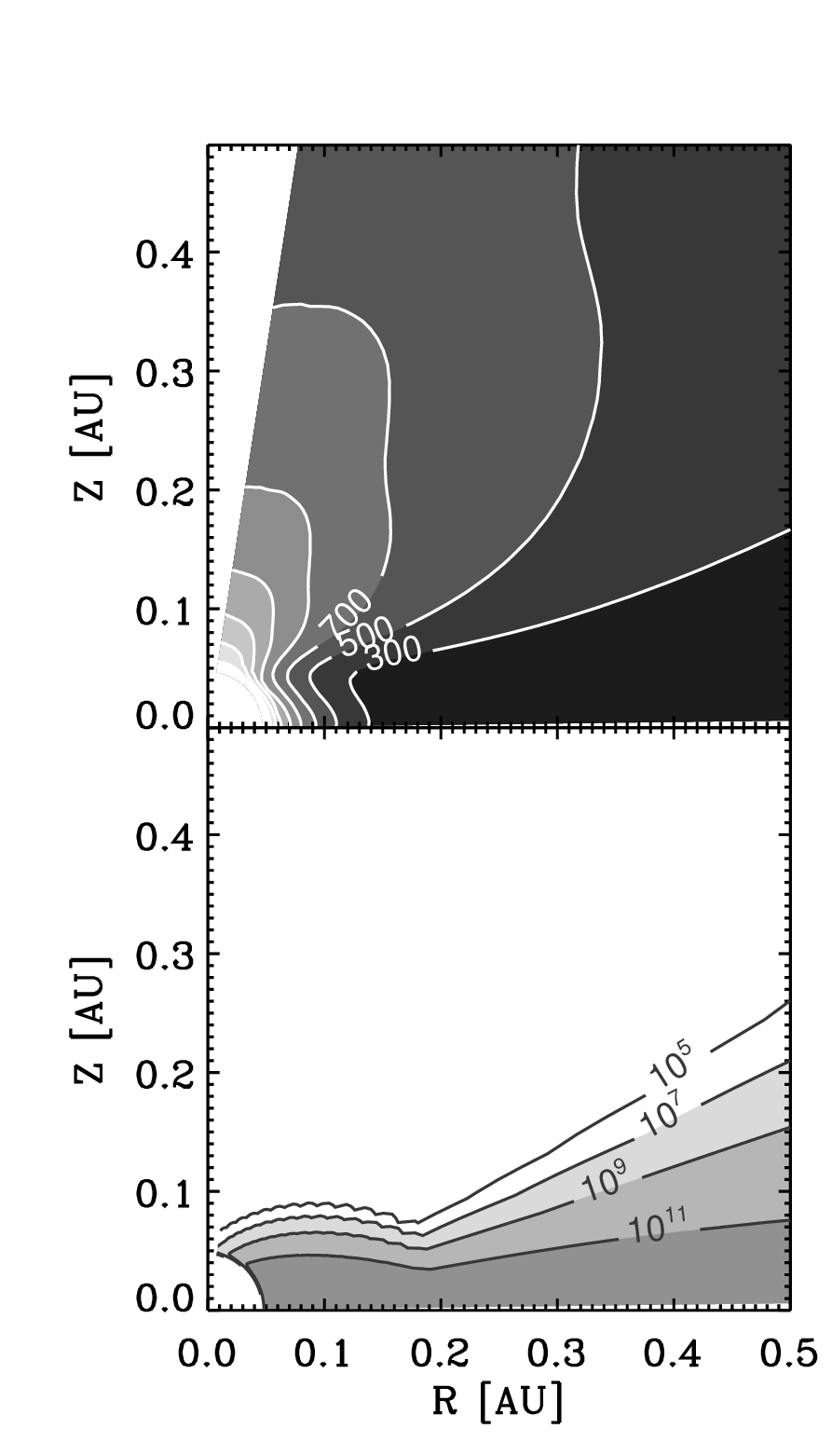}
   \caption{As in Fig. \ref{struct_EC82}, but for Ced 110 IRS 4.}
  \label{struct_CED}   
\end{figure*}

The best-fitting parameters for Ced 110 IRS 4 are given in Table \ref{ModelPars} and the temperature and density structures are 
shown in Fig. \ref{struct_CED}. 
 
The image (Fig. \ref{images_ced}) and SED (Fig. \ref{sed_ced}) of Ced 110 IRS 4 are fitted by a relatively small central disk 
with $R_{\rm disk}=35$\,AU to match the near-infrared extent of the central source. The 
central compact object in the model is not the star itself, but only light scattered in the disk surface, as is evident in the 
zoomed-in image in Fig. \ref{images_ced}. At the high inclination of the star-disk system, light scattered in the upper layers
of both the far and near side of the disk is creating the bipolar morphology seen on a 100\,AU scale in the model.
This is consistent with the fact that the observed central source is 
extended in the near-infrared ISAAC images. The size of the disk is well constrained assuming 
the flaring model disk structure is correct. The disk may be larger if it is no longer flared at radii larger 
than $R_{\rm disk}$. In other words, an outer, non-flared part of the disk 
will be shadowed from the star by the inner, flared part, and will thus not scatter any near-infrared light into the line
of sight. Near-infrared images can therefore only provide a lower bound to the physical radius of the disk. However, a non-flared
outer part of the disk will no contribute to the shadow.

The disk is surrounded by a power law envelope. The envelope is significantly more 
compact than that of CK 3 as 
indicated by the red colours and extent of the reflection nebulosity, yet the total mass of the envelope necessary to 
reproduce the data ($0.12\,M_{\odot}$) is somewhat smaller for Ced 110 IRS 4 than for CK 3.
The cavity required to produce the central rectangular dark band is assumed to be empty, apart from the central star-disk 
system. It is found that a relatively steep density profile with $\alpha=-1.5$ of the envelope is required to fit the reflection 
nebulosity. Alternatively, the envelope can be fitted by a shallower power law, which is truncated at a radius of $\sim 1500$\,AU. 
The available far-infrared data do not allow a clear distinction between these two possibilities. We choose the first option because it
conforms to other models of envelope material. A constraining property of the data is the relatively shallow 15--80\,$\mu$m 
spectral slope as measured by ISOCAM and ISOPHOT (see Fig. \ref{sed_ced}). The low 80\,$\mu$m photometric point constrains 
both luminosity and disk mass. Increasing either in the model over-predicts the far-infrared flux at 80\,$\mu$m. Decreasing 
the luminosity, but increasing the disk mass produces too little near to mid-infrared flux, and
in particular fails to reproduce the surface brightness of the reflection nebulosity. Fitting the blended IRAS and 
ISOPHOT points without regard to the data at shorter wavelengths produces a significantly higher luminosity of 1.3\,$L_{\odot}$ 
of the central object in agreement with \cite{Prusti91}, rather than the value of 0.4\,$L_{\odot}$ derived using the unblended 
photometric points. This indicates that IRS 11 accounts for the remaining 0.9\,$L_{\odot}$. The flat mid-infrared spectrum 
requires a puffed-up inner rim to the disk with a large scale height given by $H_{\rm rim}/R_{\rm rim}=0.29$. 
However, this model feature is not very important in the case of Ced 110 IRS4 in terms of the shadow, since the large-scale disk
produces most of the shadowing material, in contrast to the very tenuous disk of CK 3 where the puffed-up inner rim may produce 
most of the shadowing opacity.

The total line of sight column density toward the central disk, but excluding the disk itself, corresponds to $A_J=3.3\,$mag, 
which is consistent with the observed depth of the 9.7\,$\mu$m silicate absorption band. The predicted 3.08\,$\mu$m water ice band 
optical depth from the model is 0.7, of which 0.5 is due to the envelope. It is not possible to fit the very high $(J-H)$ colour of the 
central source of $>5\,$mag. Increasing the column density through the envelope creates a silicate band that is too deep, 
a 5--10\,$\mu$m continuum that is too low and a bipolar nebulosity that is too extinct.
This may be an indication that the dust model used is not entirely appropriate, in particular concerning the
scattering properties of the disk. The colours of the bipolar nebula and the central source can possibly be reproduced
by a somewhat flattened envelope that can produce extra extinction toward the central source, while
preserving the relatively low extinction toward each scattering lobe. However, this will still cause the mid-infrared 
extinction of the central source to be too high.

\begin{figure*}
   \centering
   \includegraphics[height=132mm,angle=90]{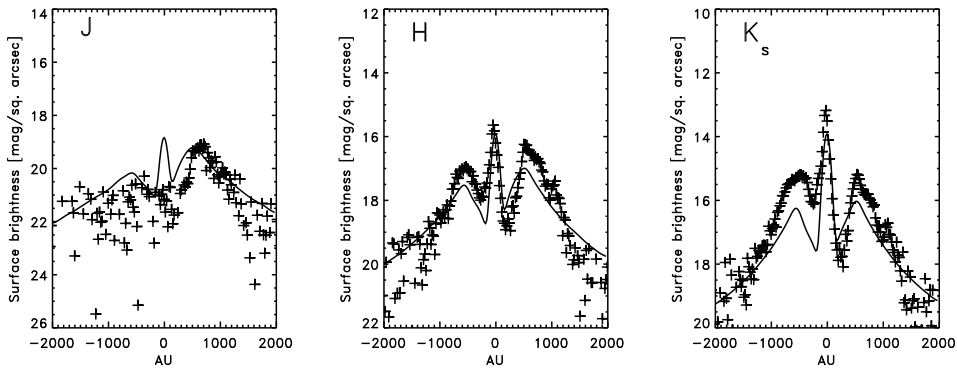}
   \caption{Cross section along the minor axis of the Ced 110 IRS 4 system for the different near-infrared wavebands. The curve 
shows the model, while the (+) symbols show the observed ISAAC surface brightness.}
   \label{cross_ced}
\end{figure*}

\begin{figure*}
\centering
   \includegraphics[width=6.5cm]{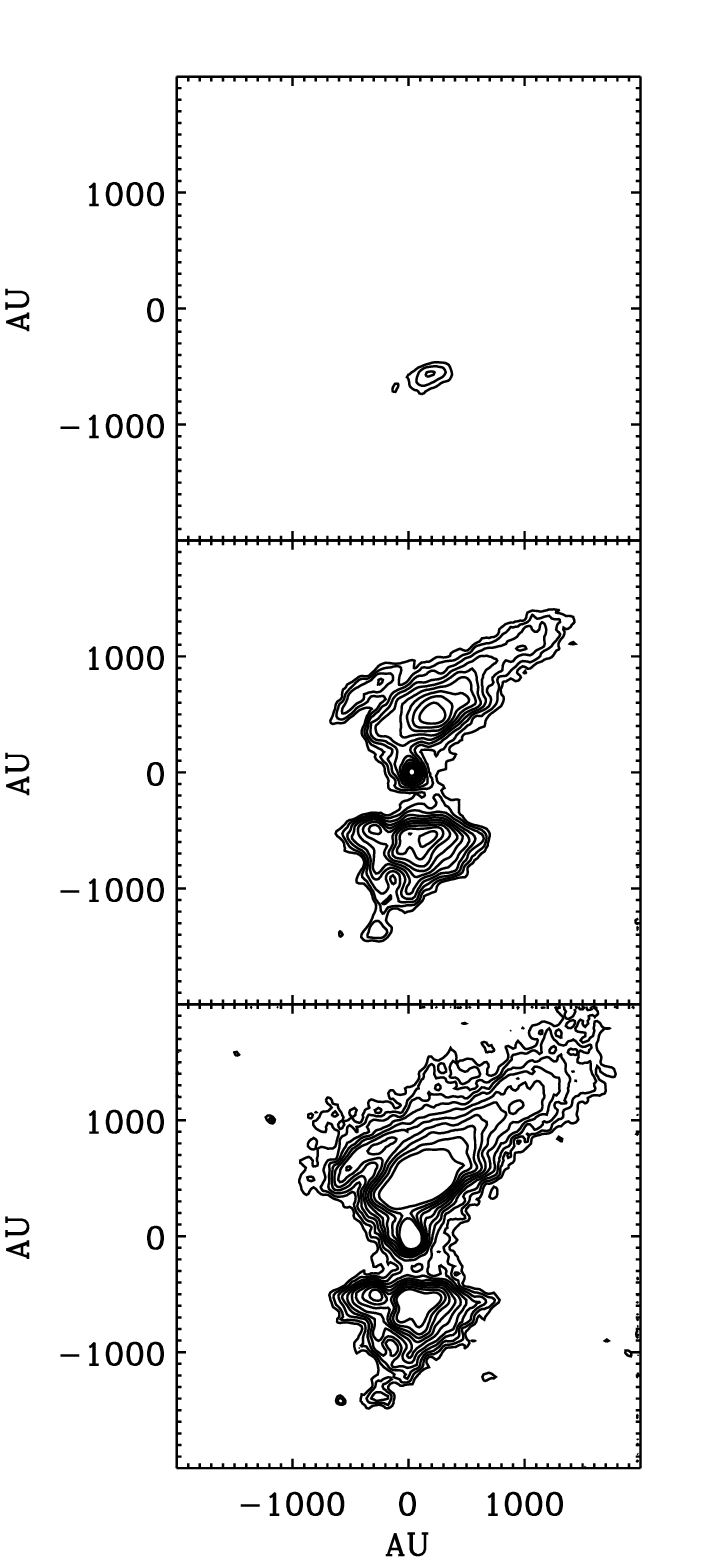}
   \includegraphics[width=6.5cm]{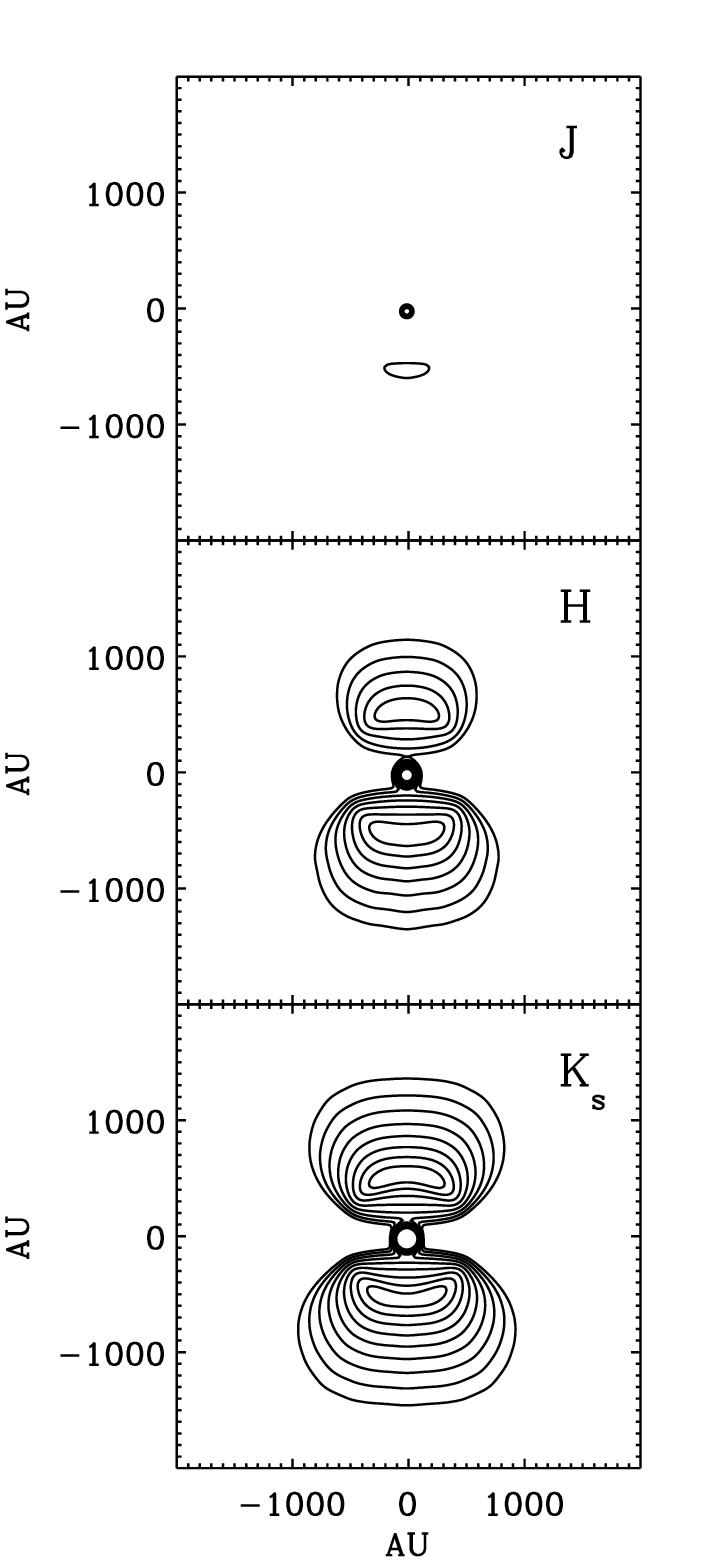}
 \caption{As Fig. \ref{EC82_contour}, but for Ced 110 IRS 4. The contours are spaced logarithmically in steps of 0.1 dex 
with the lowest contour at $\rm 1.3\times 10^{-17}$\,erg\,s$^{-1}$\,cm$^{-2}$\,Hz$^{-1}$\,sterad$^{-1}$.}
 \label{CED_contour}
\end{figure*}

Fig. \ref{cross_ced} shows a comparison of the cross section along the minor axis of the system between the model 
and observed near infrared images in terms of surface brightness. Contours of the separate near-infrared bands are 
shown in Fig. \ref{CED_contour}. It is seen that the higher reddening of the northern reflection lobe compared to the 
southern is explained by the model as a simple inclination effect. The light in the northern lobe is reflected at a 
smaller angle relative to the observer and therefore travels through a larger part of the envelope. The colour difference 
between the scattering lobes alone constrains the inclination of the system within a relatively narrow range of $70\degr\pm5$. 
This effect is important for the envelope model, because it constrains
the location of the material producing the reddening of the scattered light to the envelope rather than to a foreground component.

The size of the inner spherical cavity is determined by the position of the peaks in the reflection nebula, i.e. the width of the
dark band. However, the width of the dark band is also depending on the disk opening angle, $\Theta_{\rm disk}$. 
This means that in terms of the images, the disk opening angle and the size of the cavity are degenerate. The cavity
produces a faint circle in the model due to limb brightening. It is likely that any real cavity will neither be spherical
nor completely empty. Indeed, some diffuse emission is seen within the dark band in the ISAAC image. Small amounts of
material ($n(\rm H_2)\lesssim 10^4\,$cm$^{-3}$) present in the cavity or departures from spherical symmetry will tend to remove the 
limb brightening without affecting the shape of the shadow significantly. 

After constructing a model explaining the near-infrared morphology of Ced 110 IRS 4 as a disk shadow, it is important to 
consider whether the data can be fitted by a model without a small edge-on disk surrounding the central source. 
Specifically, the presence of a spherical cavity rather than a bipolar cavity may be difficult to explain from dynamical models.  
For instance, could the dark band be produced by a real absorbing layer of dust? It is probably possible to construct such a 
model consisting of a large (1000\,AU) torus-like structure containing the central star. However, a central hole is needed in 
any case, because a large torus or disk extending all the way to the star will completely extinct the central compact source 
in the near-infrared images. At the same time, the circumstellar material must be dense enough to prevent any scattered light 
to escape from within the dark band. Alternatively, the material producing the dark band may be located at a distance of at 
least 2000\,AU from the central star in order to suppress any reflection nebulosity. Still, the observed SED as well as the 
extended central object indicates that material is present close to the central star. We emphasise that 
the available data cannot rule out a scenario with a small disk that is not highly inclined 
and a foreground filament producing the dark band. However, the shadow scenario is in principle easy to test by obtaining 
a high resolution (0\farcs1) near-infrared image
of the central source. The shadow model shows that on 10 AU scales, the central source must appear as a compact bipolar nebula.
It is concluded that a good explanation of the data is that  a disk shadow is imprinted on 
the near-infrared reflection nebulosity. Ultimately, sensitive high resolution images in the far-infrared to millimetre 
wavelengths are needed to confirm the absence of a real flattened dusty structure, while high resolution near-infrared images 
will reveal if Ced 110 IRS 4 contains a small edge-on disk.

\section{Projection versus extinction}
\label{Selection}

\subsection{The geometry of star, disk and shadow}

\begin{figure}
   \includegraphics[width=7.0cm]{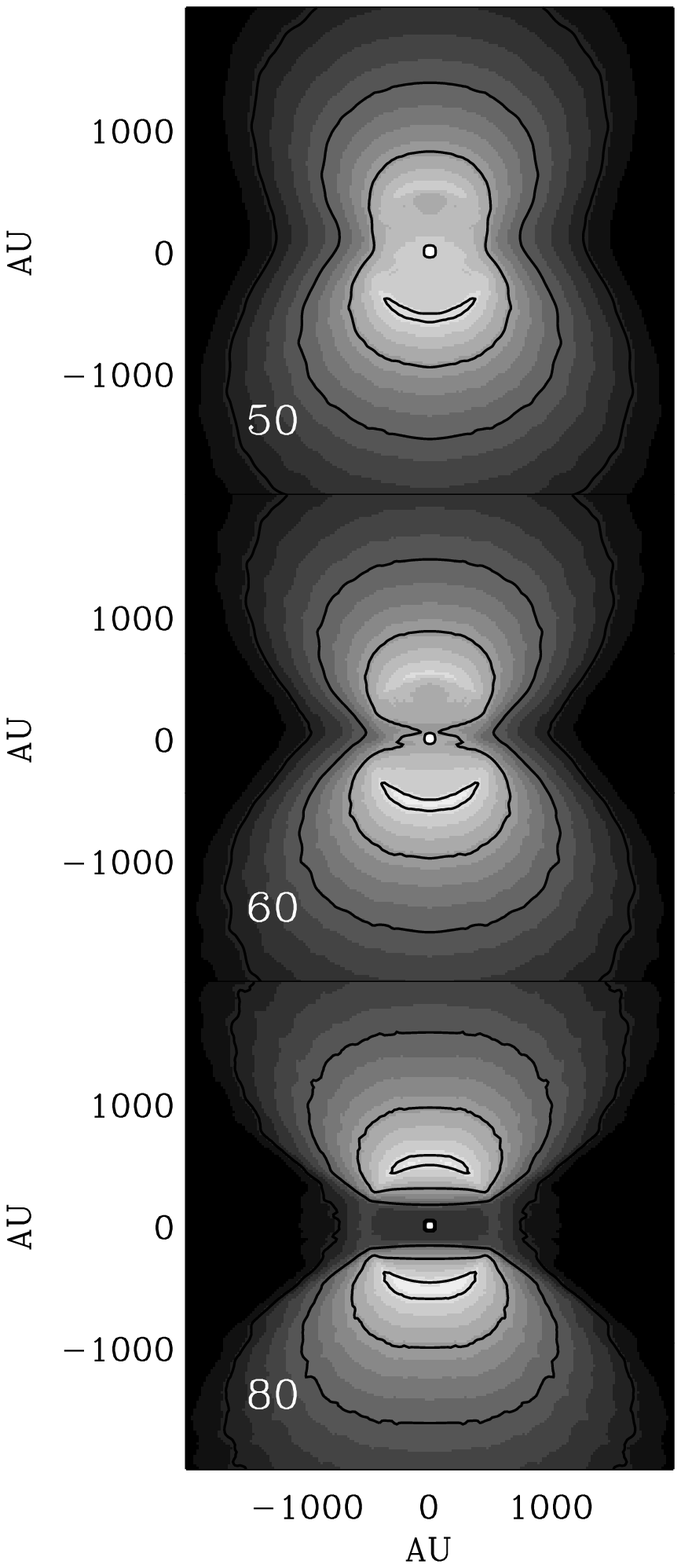}
   \caption{The model of Ced 110 IRS 4 as seen at inclinations of 50, 60 and 80\degr. The innermost contour has a value 
of $10^{-16}\,$erg\,s$^{-1}$\,cm$^{-2}$\,Hz$^{-1}$\,sterad$^{-1}$ and subsequent contours decrease in steps of 0.5 dex.}
   \label{incl_ced}
\end{figure}

The detailed geometrical relations between the central star, a flared disk and the morphology of the shadow are worth a 
closer look. Some properties are fundamental to the presence of a disk shadow. For instance, for an axisymmetric distibution 
of circumstellar material, the presence of a well-defined dark disk shadow in general means that the central star will be 
obscured by the disk. This is related to the disk opening angle, $\Theta_{\rm disk}$. For inclinations higher than 
$90\degr-\Theta_{\mathrm{disk}}$ the central compact source consists of light reflected off the surface of the disk, the 
central star being obscured by the disk itself. At relatively low inclinations, disk shadows still affect the observed 
near-infrared morphology without showing clearly defined dark bands. Fig. \ref{incl_ced} shows the effects of different 
inclination angles of Ced 110 IRS 4. At inclinations $i<90\degr-\Theta_{\mathrm{disk}}$ no dark lane is seen, but the 
nebulosity is still significantly elongated. This means that disk shadows can be mistaken for outflow cavities even at 
this inclination, but can be distinguished by the presence of a stellar source, i.e. the central star can be seen directly 
over the edge of the disk. If the shadow is cast on a backdrop `screen' located entirely behind the source, a dark band 
will appear also for $i<90\degr-\Theta_{\mathrm{disk}}$  and the central compact source will appear outside the shadow 
(see Fig. \ref{screen_cartoon}). Again, if the source is located outside the shadow for a backdrop screen, the central 
star must become visible and will outshine the scattered emission from the physical disk. The location of the central 
source relative to the shadow as seen in a near-infrared image is thus intimately linked to the visibility of the central star. 

The strongly inter-related properties of visibility of the central star, near-infrared morphology and to a lesser degree 
the SED of the source means that in particular the inclination of the model system must be tightly constrained.

\subsection{Selection criteria for disk shadows}

A number of properties common to projected disk shadows can be identified from the models and compared to
extinction from large disks in order to better classify any observed dark lane signatures in infrared images. 
In this section, we will attempt to identify useful selection criteria for disk shadows in the absence of sensitive 
imaging at long wavelengths.

The most important difference between dark lanes due to extinction from a large disk and the dark lanes due to projection 
is that the former are expected to be entirely dark while the latter will still have a bright source in the center of the 
system. Such a morphology is difficult to produce with a standard large disk model. For a very tenuous disk, the central 
star may of course be seen, such as is the case for CK 3. A very tenuous disk will produce a wedge-shaped morphology similar 
to a `scenario 1' shadow, as seen in Fig. \ref{images_EC82}. However, the scattering lobes of a large tenuous disk will be 
flattened in the absence of a spherical envelope. Typically, the bright source within the dark lane is indirect stellar 
light which is being scattered on the disk surface into the line of sight. Much of this light may still be extincted by 
the envelope, such that the blueing effect of scattering is partly compensated by reddening due to extinction by the envelope 
material along the line of sight to the central source.

A disk shadow will for `scenario 1' be characterised by a symmetric wedge-shaped band which
covers the entire extent of the nebula illuminated by the central star. Typically, but not necessarily, 
the shadow will have an angular size which is much larger than what is expected from a disk. 
Clearly, there should be no unambiguous evidence of a physical disk {\it of the same size}. Conversely, a smaller edge-on 
disk must be present to cast the shadow. Thus, it should be possible to resolve the central object into an edge-on disk 
with a dark lane in high-resolution near-infrared images. Alternatively, since the light of the central compact source 
in the image consists exclusively of light scattered into the line of sight by grains in the disk surface layers
(the direct starlight being totally obscured), it must be highly polarised to a degree that is easily observable. High 
resolution imaging or polarimetry of the central sources are therefore reliable methods to determine if a 
dark band is a disk shadow or not. For inclinations significantly higher than $90\degr-\Theta_{\mathrm{disk}}$, the edge 
of the shadow will be straight and sharp, but for inclinations of $\lesssim 90\degr-\Theta_{\mathrm{disk}}$, some curvature 
may be seen. For `scenario 2', i.e. the screen shadow scenario, in which the central source is surrounded by a cavity, 
a dark band should be seen but with a central near-infrared compact source in the center. The presented model never 
produces a disk shadow without the central source being at least as bright as the reflecting material. 

\begin{table*}
\footnotesize
\centering
\begin{flushleft}
\caption{Selection criteria for disk shadow candidates}
\begin{tabular}{lll}
\hline 
\hline 
 & Disk shadow & Large edge-on disk  \\
\hline
Sharp edges of the dark band &most disk shadows&most large disks\\
`Wedge-like shape'&`Scenario 1' disk shadows&only for very tenuous disks\\
Extremely large size in the NIR ($\rm >2000\,AU$)&most disk shadows&rarely\\
Source in the center&`Scenario 1+2b' disk shadows&only for very tenuous disks\\
Variability of the dark band&possible for all types of disk shadow&very unlikely\\
Central source is bipolar on $\sim 1\arcsec$ scales &for $i>90\degr-\Theta_{\mathrm{disk}}$&only tenuous large 
disks have a central source\\ 
\hline
\end{tabular}
\label{seltable}
\end{flushleft}
\end{table*}

In some cases, significant shadowing can be produced by disk structures within a few AU of the star, such as may be the case for CK 3. 
Shadows produced by material located close to the central star may be highly variable on timescales as short as days. Conversely, large
edge-on disks producing dark bands will have much longer timescales for variation. Thus, a variable large dark band is a clear signature
of a disk shadow.
The selection criteria for disk shadows are summarised in Table \ref{seltable} and compared to bona-fide large disks.

\section{Concluding remarks}
\label{Discussion}
We have argued that circumstellar disks may project large shadows onto any nebulosity 
illuminated by the central star, regardless of whether the scattering material is located around the
star, in the background or in the foreground. The disk shadows will be most clearly recognisable if the
disk is close to edge-on, but dark bands will also be present around disks with inclinations down to $\sim 70\degr$, and
significant bipolar elongations of the nebulosity may be seen to much lower inclinations.
We have presented two specific cases that fit well the selection criteria for each of the two
scenarios and we have shown that all available spectroscopic and imaging data can be fitted by disk shadow models. 
Both candidates have very conspicuous morphologies of the near-infrared reflection nebulosity. However,
many disk shadows may exist that are more difficult to identify due to an asymmetric distribution
of circumstellar material. The main objective has
been to argue that a range of different observed near-infrared morphologies can be explained as large shadows cast by circumstellar 
disks of average size ($R_{\rm disk}\sim 100$\,AU, $M_{\rm disk}\sim 10^{-6}-10^{-2}\,M_{\odot}$). The fits to the
data are not necessarily unique in every respect, although inclination and infrared extinction through the disk are 
well-constrained. Some parameters such as the total disk mass may change significantly if some model assumptions are changed. 
In particular, the effect of different grain size distributions is not explored and the amount of settling of large dust grains
may play a role for the derived parameters. Indeed, the very low dust mass in small grains derived for CK 3 may be taken as 
evidence for significant dust growth and settling in this source. However, if the mass estimates are correct and there
is no `missing mass' hidden in large grains, these disks may have interesting dynamic properties. 
The disk mass in small grains as given by both models is 4-5 orders of magnitude less than the inferred envelope masses. This means
that any interaction between the disks and envelopes may be different than typically assumed. A direct interferometric measurement
of the disk mass, possibly with ALMA is necessary to resolve the issue.

For the two disk shadow candidates discussed here, the near-infrared images provide information 
which is highly complementary to fitting a model to the SED only. The presence of a disk shadow
therefore may break some of the degeneracies that are inherent in SED fitting. For instance, the disk
shadow of CK 3 clearly shows that the system is highly inclined, while the SED alone suggests that the
system is viewed close to face-on due to the emission features in the spectrum. Therefore, disk shadows unambiguously identify
edge-on disk systems without the need for very high spatial resolution imaging.  Furthermore, if the
variability of the disk shadow of CK 3, suggested by \cite{Sogawa97}, is real, the structure and dynamics of
the inner rim of the disk can be constrained in a unique way. The morphology of the shadow may in some
cases constrain the distribution of envelope material. This is central to the division of disk shadows in `scenario 1' shadows 
projected {\it into} and `scenario 2' shadows projected {\it onto} the envelope material.  Only a few identified edge-on
disks surrounding low-mass stars are known and significantly more may be added using the signatures of disk shadows.
In other words, a disk shadow source may provide modelling constraints for an unresolved disk similar to those obtained 
for resolved edge-on disks such as the Butterfly Nebula \citep{Wolf03} or HH 30 \citep{Wood02}.

Also, edge-on disks are important for studying disk material which can only be observed in absorption, such as ices \citep{Thi02,Pontoppidan04}. The possibility of identifying edge-on disks using disk shadows is interesting for
observing ices in disks because such objects may be too tenuous and small to
be detected through other means. Ices are expected to be abundant in the cold mid-planes of circumstellar disks. However, the very high optical depth through the mid-planes of many disks prevents the formation of ice absorption lines because
the infrared continuum of the inner parts of the disk is completely obscured. Disks with a relatively low extinction through the mid-plane such as CK 3 may be excellent candidates for ice observations, yet they may be difficult to identify as edge-on disks in the absence of a disk shadow. As it turns out, the CK 3 disk optical depth along the mid-plane is so low that few ices are seen, but other disk shadow candidates may be better suited for searching for ices in disks. The disadvantage of disk shadows for ice observations
is that the presence of a shadow dictates the presence of envelope material, which may contaminate the signal from disk ices.

Disk shadow scenarios may also be used to explain peculiar near-infrared morphologies.
It is surprising how well a spherically symmetric envelope fits the reflection nebulosity of the two sources discussed in this paper when disk shadowing is taken into account. Specifically, no {\it bipolar} cavity is required for Ced 110 IRS 4, although it does require an inner {\it spherical} cavity to produce a dark band with parallel edges. However, it is important to note that many bipolar nebulosities still require bipolar cavities, especially for small opening angles of the bipolar nebulosity. How an inner spherical cavity is created remains an open question.

Apart from the disk shadow candidates modelled here, CK 3 of the Serpens Reflection Nebula and Ced 110 IRS 4, other examples may 
include some (but not all) of the bipolar nebulae in Taurus imaged by \cite{Padgett99}. In particular, IRAS 04248+2612 and CoKu Tau/1 
of this sample resemble `scenario 1' disk shadows. Other sources also show a near-infrared morphology very similar to that of Ced 110 
IRS 4, i.e. a rectangular dark lane in a bipolar nebulosity with a compact source in the center. For instance, judging from the 
imaging of \cite{Brandner00}, LFAM 26 (CRBR 2403.7-2948) is a good candidate for a `scenario 2' disk shadow. 
The `Flying Ghost Nebula' \citep{Boulard95} shows the wedge-like morphology of a `scenario 1' shadow. In \cite{McCaughrean02}, 
a deep H$_2$ image observed with VLT-ISAAC of the region around the protostellar outflow HH 212, clearly shows the presence
of a `scenario 2b' disk shadow candidate, exhibiting a reflection nebulosity intersected by a broad, dark band and a compact
source in the center of the dark band. Finally, a recently 
discovered peculiar massive YSO in M17,  suggested by \cite{Chini04} to be an extremely large accretion disk of at least 
100\,$M_{\odot}$, shows many of the characteristics of a projected disk shadow of a much smaller disk. In particular, the object 
has a symmetric, wedge-shaped morphology extending over more than 30\,000\,AU and a central bright source. As such it bears strong 
resemblance to CK 3 and the Serpens reflection nebula. It is shown here that it is easy to produce such a morphology in terms of a 
disk shadow. The authors suggest the disk in M17 is seen in silhouette against the emission of the background
HII region. In contrast, a shadow scenario requires the central star to be the local ionising source and a small disk to 
shadow ionising photons, rather than scattered near-infrared photons. 
This will produce a wedge-shaped disk of neutral gas that will appear dark in relation to the surrounding
ionised gas. Similarly, a disk shadow surrounding a UV source 
may also produce a large molecular disk, not maintained by rotation, but rather by the lack of dissociating photons 
in the plane of the shadowing disk.

\begin{acknowledgements}This research is supported by a PhD grant from the Netherlands
 Research School for Astronomy (NOVA).  Support for this work, part of the Spitzer Space Telescope Legacy Science Program, was provided by NASA through Contract Numbers 1224608 and 1230780 issued by the Jet Propulsion Laboratory, California Institute
of Technology under NASA contract 1407. Jackie Kessler-Silacci is thanked for providing a cleaned version of the Spitzer spectrum of CK 3.
The authors are grateful to Ewine van Dishoeck and Geoff Blake for careful reading of the manuscript.
This research was supported by the European Research Training Network "The Origin of Planetary Systems" 
(PLANETS, contract number HPRN-CT-2002-00308).
 \end{acknowledgements}

\bibliographystyle{aa}
\bibliography{aa2059}

\end{document}